\documentclass[preprint]{aastex}
\usepackage{psfig}
 
\newcommand{\av}{\ensuremath{A_V}}

\newcommand{\etal}{et\,al.}

\newcommand{\ngc}{NGC\,7538 IRS9}

\newcommand{\sgra}{Sgr\,A*}
\newcommand{\sgraw}{Sgr\,A\,W-IRS3}
\newcommand{\ammonia}{NH$_3$}
\newcommand{\formald}{H$_2$CO}
\newcommand{\methane}{CH$_4$}
\newcommand{\methanol}{CH$_3$OH}
\newcommand{\methylene}{-CH$_2$-}
\newcommand{\methyl}{-CH$_3$}

\newcommand{\water}{H\ensuremath{_2}O}

\begin{document}

\title{The Composition and Distribution of Dust Along the Line of Sight Towards
the Galactic Center\normalsize{\altaffilmark{1}\altaffiltext{1}{Based on
observations  made with ISO, an ESA project with instruments funded by ESA
member states (especially the PI countries: France, Germany, the Netherlands,
and the United Kingdom) and with the participation of ISAS and NASA.}}}

\author{J.E.\ Chiar\altaffilmark{2}, A.G.G.M. Tielens\altaffilmark{3}, D.C.B.
Whittet\altaffilmark{4}, W.A. Schutte\altaffilmark{5}, A.C.A.
Boogert\altaffilmark{3,6}, D. Lutz\altaffilmark{7},
E.F. van Dishoeck\altaffilmark{5}, M.P. Bernstein\altaffilmark{8}}
\altaffiltext{2}{NASA/Ames Research Center, Mail Stop 245-3, Moffett
Field, CA 94035 and SETI Institute, Mountain View, CA
94043. \texttt{mailto:chiar@misty.arc.nasa.gov}}
\altaffiltext{3}{Kapteyn
Astronomical Institute, P.O. Box 800, 9700 AV Groningen, The Netherlands}
\altaffiltext{4}{Rensselaer Polytechnic Institute, Department of Physics,
Applied Physics and Astronomy, Troy, NY 12180}
\altaffiltext{5}{Leiden Observatory, P.O. Box 9513, 2300 RA Leiden, The
Netherlands}
\altaffiltext{6}{present address: California Institute of Technology, Downs
Laboratory of Physics, Mail Code 320-47, Pasadena, CA 91125}
\altaffiltext{7}{Max-Planck-Institut f\"ur extraterrestrische Physik, Postfach
1603, 85740 Garching, Germany}
\altaffiltext{8}{NASA/Ames Research Center, Mail Stop 245-6, Moffett Field, CA
94035 and SETI Institute, Mountain View, CA 94043}

\begin{abstract}
\small
We discuss the composition of dust and ice along the line of  sight to the
Galactic Center (GC) based on analysis of  mid-infrared spectra (2.4--13
$\mu$m) from the Short  Wavelength Spectrometer on the Infrared Space
Observatory  (ISO).  We have analyzed dust absorption features arising in
the  molecular cloud material and the diffuse interstellar medium  along
the lines of sight toward Sagittarius A* and the  Quintuplet sources, GCS3
and GCS4. It is evident from the  depth of the 3.0 $\mu$m H$_2$O and the
4.27 $\mu$m CO$_2$  ice features that there is more molecular cloud
material  along the line of sight toward Sgr A* than GCS3 and 4. In  fact,
Sgr A* has a rich infrared ice spectrum with evidence  for the presence of
solid \methane, NH$_3$, and  possibly HCOOH. 

Hydrocarbon dust in the diffuse interstellar medium along  the line of
sight to the GC is characterized by absorption  features centered at 3.4
$\mu$m, 6.85 $\mu$m, and 7.25  $\mu$m. Ground-based studies have identified
the 3.4 $\mu$m  feature with aliphatic hydrocarbons, and ISO has given us 
the first meaningful observations of the corresponding modes  at longer
wavelengths. The integrated strengths of these  three features suggest
that hydrogenated  amorphous carbon is their carrier. We attribute an 
absorption feature centered at 3.28 $\mu$m in the GCS3  spectrum to the
C-H stretch in aromatic hydrocarbons.  This feature is not detected, and
its C-C stretch counterpart appears to be weaker, in the \sgra\ spectrum. 
One  of the key questions which now arises is whether   aromatics are a
widespread component of the diffuse  interstellar medium, analogous to
aliphatic  hydrocarbons. 

\end{abstract}
\normalsize


\section{Introduction}\label{intro}
Bright infrared (IR) sources combined with 30 magnitudes of visual
extinction have made the Galactic Center (GC) an attractive line of sight
for the  study of interstellar dust composition. Absorption features in
the spectra of GC IR sources have been used extensively to characterize
interstellar grain material, including  silicates
\nocite{becklin_etal78I,roche_aitken85} (Becklin \etal\ 1978; Roche \&
Aitken 1985) and  hydrocarbons
\nocite{willner_etal79,mcfadzean_etal89,sandford_etal91,pendleton_etal94}
(Willner \etal\ 1979; McFadzean \etal\ 1989; Sandford \etal\ 1991;
Pendleton \etal\ 1994). Previous observations of the  GC suggested the
presence of absorptions expected in icy grain material, most notably the
3$\mu$m band of \water-ice (e.g., McFadzean \etal\ 1989), but until
recently, the explanation that one or more molecular clouds obscure the GC
and contribute to its total extinction had not been completely accepted
since the 3$\mu$m feature has a substantially different profile than that
observed in dense cloud sources (Tielens \etal\ 1996; Wada \etal\ 1991).
The presence of molecular cloud material in the GC was confirmed via
detection  of solid CO$_2$ feature by the Short Wavelength Spectrometer
(SWS) on the Infrared Space Observatory (ISO)
\nocite{gerakines_etal99,degraauw_etal96b,lutz_etal96} (Gerakines \etal\
1999; de Graauw \etal\ 1996b; Lutz \etal\ 1996). 

Solid organic material is a well-established component of the diffuse
interstellar medium.    The complex 3.4 \micron\ feature,
ubiquitous along sightlines which sample the diffuse ISM in our Galaxy and
other galaxies \nocite{pendleton_etal94,bridger_wright_geballe93}(e.g.
Pendleton \etal\ 1994, Bridger, Wright, \& Geballe 1993),  is identified
with short-chained aliphatic hydrocarbons with a methylene/methyl
(-CH$_2$-/-CH$_3$) group ratio between 2--2.5 (Sandford \etal\ 1991;
Pendleton \etal\ 1994). The CH deformation mode at  6.8 \micron\ was
detected in KAO data (Tielens \etal\ 1996), although the weaker 7.3
\micron\ band could not be confirmed at the time.  One model proposes 
that organic refractory matter in the diffuse ISM is the product of
energetic processing of simple ices (such as CO, \water, \methanol) in
dense clouds \nocite{allamandola_sandford_valero88,greenberg_etal95,
strazzulla_castornia_palumbo95,moore_ferrante_nuth96} (e.g., Allamandola,
Sandford, \& Valero 1988;  Greenberg \etal\ 1995; Strazzulla, Castornia,
\& Palumbo 1995; Moore, Ferrante, \& Nuth 1996). However, the discovery of
spectroscopic similarity between diffuse interstellar and circumstellar
organics has prompted a reassessment of the origin of organic  matter in
the ISM \nocite{lequeux_muizon90,chiar_etal98crl}(Lequeux \& Jourdain de
Muizon 1990; Chiar \etal\ 1998b).  The striking similarity of the
circumstellar and interstellar 3.4 \micron\ features indicates that at
least  some  of the organic carriers which lead to the 3.4 \micron\
absorption in the  diffuse ISM originate as stardust (Chiar \etal\ 1998b). 
The case for production of the 3.4 \micron\ band carrier in dense clouds
and thus its presence in organic  refractory mantles is further weakened
by the finding of \nocite{adamson_etal99} Adamson \etal\ (1999) that,
unlike the 9.7 \micron\ silicate feature \nocite{aitken_etal86}(e.g.,
Aitken \etal\ 1986),  the 3.4 \micron\ feature along the line of sight
toward \sgra\ (IRS7) is not polarized.   Hence,  the  3.4 \micron\ band is
not produced by a carrier residing in a mantle on a silicate core, but
rather by very small (unaligned) grains.  Nevertheless, energetic
processing of ices in dense clouds may still play some role in the 
production of interstellar organics since diffuse ISM dust cycles into
dense clouds fairly rapidly \nocite{mckee89,jones_etal94}($\sim
3\times10^7$ years; McKee 1989; Jones \etal\ 1994).  

Hydrogenated Amorphous Carbon (HAC) is  a possible candidate for the
organic component of diffuse interstellar dust (Tielens \etal\ 1996;
Pendleton \etal\ 1994), and several laboratory groups are attempting to
produce HAC analogs with a -CH$_2$-/-CH$_3$ ratio which matches the
observed feature \nocite{furton_etal99,duley_etal98}(e.g., Furton \etal\
1999; Duley \etal\ 1998).  Previous comparisons of HACs with the
interstellar 3.4 \micron\ feature have shown that many  HACs produced in
the laboratory are lacking in -CH$_3$ groups relative to -CH$_2$- groups,  
providing an unsatisfactory fit to the
observed feature (e.g., Pendleton \etal\ 1994).  Recently, Duley \etal\
(1998) have shown that the 3.38 \micron\ -CH$_3$ subfeature is
strengthened at lower temperatures (77\,K), and the resulting HAC with
-CH$_2$-/-CH$_3<1$, provides a better match to the 3.4 \micron\
interstellar feature, compared with higher temperature (e.g., 300\,K)
HACs.    

Recently, a feature near 6.2 \micron\ was discovered along sightlines
toward the GC Quintuplet sources and several dusty late-type WC Wolf-Rayet
stars, and has been attributed to the aromatic C-C stretch in polycyclic
aromatic hydrocarbon molecules (PAH) in the diffuse ISM along the line of
sight (Schutte \etal\ 1998).   PAHs are a certain component of the
carbonaceous material in the Galaxy, as shown by the dominance and
ubiquitous presence of the  family of PAH emission features at 3.3, 6.2,
7.7, 8.6, and 11.3 \micron\  in diffuse emission in the Galactic disk
\nocite{giard_etal88,mattila_etal96,ristorcelli_etal94}(Giard \etal\
1988;  Ristorcelli \etal\ 1994; Mattila \etal\ 1996).   Therefore, the
detection of the aromatic C-C stretch at 6.2 \micron\ in diffuse ISM
sightlines is not surprising; its carrier may be in the form of  gaseous
polycyclic aromatic hydrocarbons (PAHs) or PAH clusters with sizes up to a
few thousand C atoms \nocite{schutte_etal98}(Schutte \etal\ 1998).  The
corresponding C-H stretch at $\sim3.3$ \micron\ was not detected in
absorption in the diffuse ISM, although high signal-to-noise data,
necessary to detect this weak feature, were not available for many
sightlines.  However, an absorption feature at 3.25 $\micron$ is detected
in ground-based data in five heavily extincted molecular cloud
sightlines \nocite{brooke_sellgren_geballe99,sellgren_etal95}(Sellgren
\etal\ 1995; Brooke, Sellgren, \& Geballe 1999).

In this paper we discuss the composition of the  hydrocarbon dust
component as well as the molecular cloud ices, along the line of sight to
the Galactic Center based on 2.5--13 \micron\ spectra from the Infrared
Space Observatory's Short Wavelength Spectrometer (ISO-SWS).  ISO-SWS had
revealed the previously elusive 2.95 \micron\ \ammonia\ feature in dense
cloud material along the GC line-of-sight.  These observations have also
provided us with the first meaningful observations of hydrocarbon modes at
long wavelengths (5--8\micron) corresponding to the well-studied 3.4
\micron\ absorption feature, associated with the diffuse ISM, allowing us
to narrow down the list of possible hydrocarbon candidates
(\S\ref{aliphatics}).  We also present the first detection of aromatic
absorption, at 3.28 \micron, unassociated with molecular cloud dust.  
This paper is organized as follows:  In
\S\ref{sources} we briefly discuss what is known about the nature of the
sources, and in \S\ref{obs} we explain the data reduction, and present the
ISO-SWS spectra of 3 sources in the Galactic Center.  In the following
sections, we discuss the nature of the molecular cloud IR absorption 
features (\S\ref{molecular}) and hydrocarbon absorption features
(\S\ref{diffuse}) along the line of sight.   Finally, we  discuss possible
scenarios for the origin of hydrocarbons in the GC region
(\S\ref{discussion}) and summarize our results (\S\ref{summary}).

\section{The Nature of the Sources}\label{sources}

In this paper we analyze spectra from ISO-SWS for three sightlines toward
the   Galactic Center: \sgra (and \sgraw), GCS3, and GCS4. The IR sources
contained in the ISO beam along the line of sight toward \sgra\ and
\sgraw, suffer visual extinction  $\sim31$ magnitudes
\nocite{rieke_etal89}(Rieke \etal\ 1989, and references therein), and 
contain M giants and supergiants, and \ion{H}{2} regions
\nocite{lebofsky_etal82,sellgren_etal87,wollman_smith_larson82,eckart_etal95}
(Eckart \etal\ 1995; Sellgren \etal\ 1987; Lebofsky \etal\ 1982; Wollman,
Smith, \& Larson 1982).  GCS3 and 4, located near the GC Radio Arc,
approximately 14' NE of the Galactic Center, corresponding to a projected
distance of about 30 pc (assuming a distance of 8.2 kpc), were first
discovered in the polarimetric study by \nocite{kobayashi_etal83}
Kobayashi \etal\ (1983).  Visual extinction estimates are similar to that
for the GC, \av$\sim29$ magnitudes (Figer \etal\ 1999 and references
therein).  Detailed imaging  of the region  revealed that GCS3 and 4
consist of a compact group of at least 5 IR sources, which have since
become known as the Infrared Quintuplet
\nocite{nagata_etal90,okuda_etal90}(Nagata \etal\ 1990; Okuda \etal\
1990), although more recent imaging and spectroscopic investigations 
\nocite{moneti_etal92,moneti_etal94} (Moneti \etal\ 1992, 1994) have shown
that this region contains several bright and many faint IR sources.  The
nature of these objects is still a question of debate: several studies
have suggested that the Quintuplet cluster consists of  young stars
surrounded by thick shells or cocoons of hot dust
\nocite{moneti_etal94}(e.g., Moneti \etal\ 1994).   A very recent
photometric study has proposed that these stars are actually massive
late-type dusty WC Wolf-Rayet stars \nocite{figer_mclean_morris99} (Figer,
McLean, \& Morris 1999).  

\section{Observations and Data Reduction}\label{obs}

Infrared spectra of Sagittarius A* (Lutz \etal\ 1996), Sagittarius A
W-IRS3, and Quintuplet (AFGL 2004) sources GCS3 and GCS4 were obtained
with ISO-SWS in 1996 and 1997.  The reader is referred to
\nocite{degraauw_etal96a} de Graauw \etal\ (1996a) for a more detailed
description of SWS and its capabilities.  Observational details and
coordinates are listed in Table~1.  The ISO-SWS $14\arcsec\times20\arcsec$
beam was centered on the coordinates listed in  Table~1; the \sgra\ and
\sgraw\ positions are only several arcseconds apart, and therefore probe
similar lines of sight in the relatively large ISO beam.  The observations
of GCS3 were centered on GCS3-I and included all four GCS3 objects; GCS4,
located some 15\arcsec\ away from GCS3-I probes a unique sightline in this
region.  Data for GCS3 and GCS4 have been presented in Schutte \etal\
1998; we have re-reduced these data using the most up-to-date calibration
files (see below).  Astronomical Observing Template 1 (AOT1), speed 4, was
used to obtain data for SgrA* in the 2-45 \micron\ spectral region (Lutz
et al. 1996), giving resolving power $R\sim800-1700$ depending on the
wavelength region.  AOT1 speed 3 data were obtained for the GCS3 and 4 in
the full 2--45 \micron\ spectral region with $R\sim480-1020$.  Higher
resolution data were obtained for GCS3 with AOT6 in the 3.1--3.5
($R\sim2000$) and 5.3--7.0 \micron\ ($R\sim1350$) spectral regions, and
for \sgraw\ in the 7.4--8.0 \micron\ ($R\sim1700$) spectral region.  

\begin{figure}
\centerline{\psfig{figure=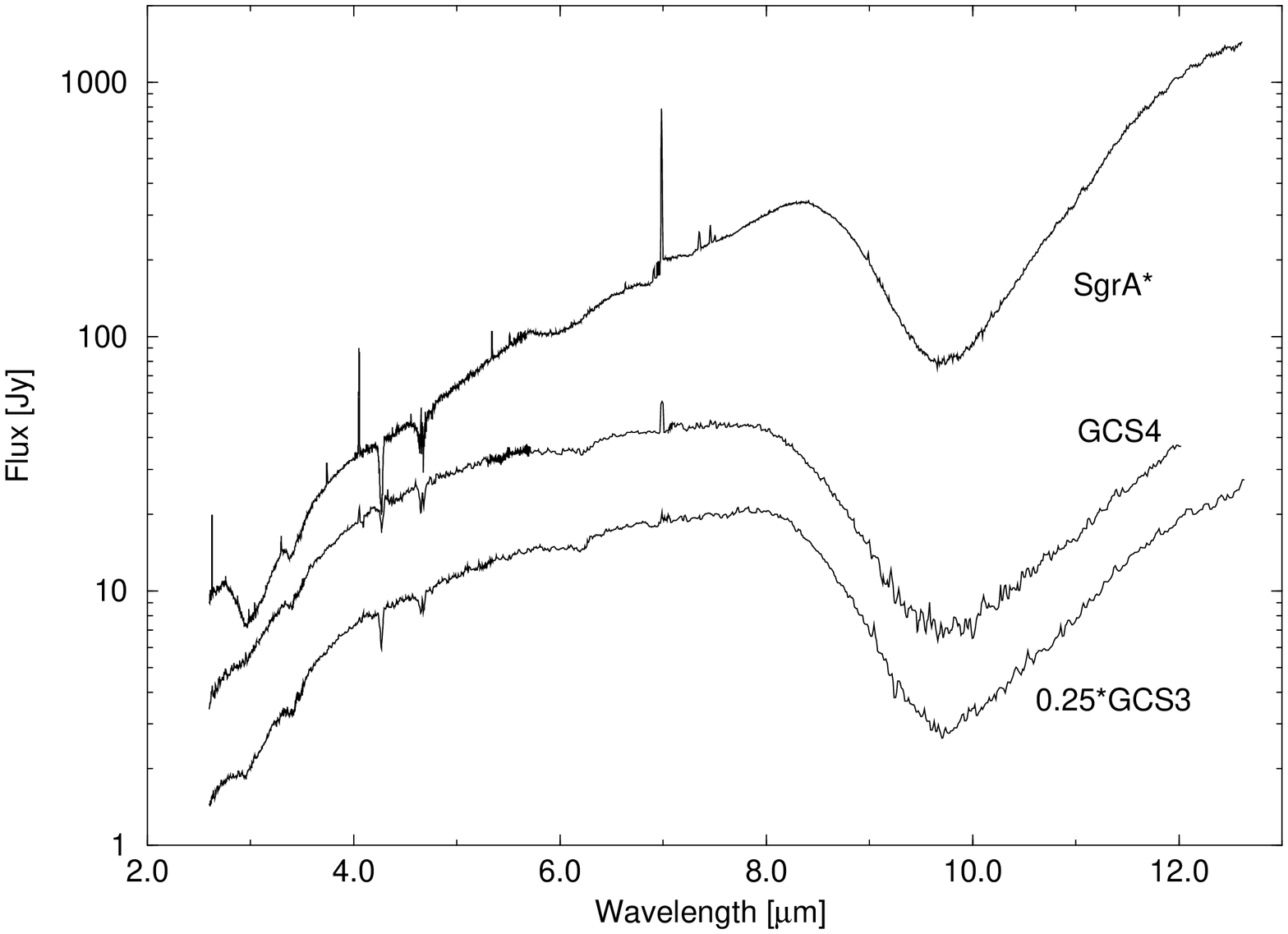,height=3in,angle=-90}}
\figcaption{ISO-SWS spectra from 2.5--13 \micron\ of SgrA* (AOT6), GCS4 (AOT1)
and GCS3 (AOT1).  The GCS3 spectrum has been divided by 4 for presentation
purposes.
\label{fig:all_full}}
\end{figure}

In the present paper, we discuss the 2.5--13 \micron\ spectral region
which contains the strongest absorption features of hydrocarbons and ices,
except for the 15.3 \micron\ CO$_2$ band which has been discussed
extensively by Gerakines \etal\ (1999) and references therein.  
\nocite{moneti_etal92} The data were reduced using the SWS Interactive
Analysis package (de Graauw \etal\ 1996a) and the Observers SWS
Interactive Analysis Package (OSIA).  Version 7 of the SWS calibration
files were used.   The flux-calibrated spectra from 2.5--13 \micron\ are
shown in Fig.~\ref{fig:all_full}.  Optical depth spectra were determined
by fitting local low order polynomial continua as described in the
following sections.  No evidence for broad molecular emission features
(e.g., due to PAHs) was
apparent in the spectra, thus continua were chosen assuming only
the presence of absorption. Optical depths of the dust absorption
features are listed in Table~2.

\section{The Molecular Cloud Features}\label{molecular}

\begin{figure}
\centerline{\psfig{figure=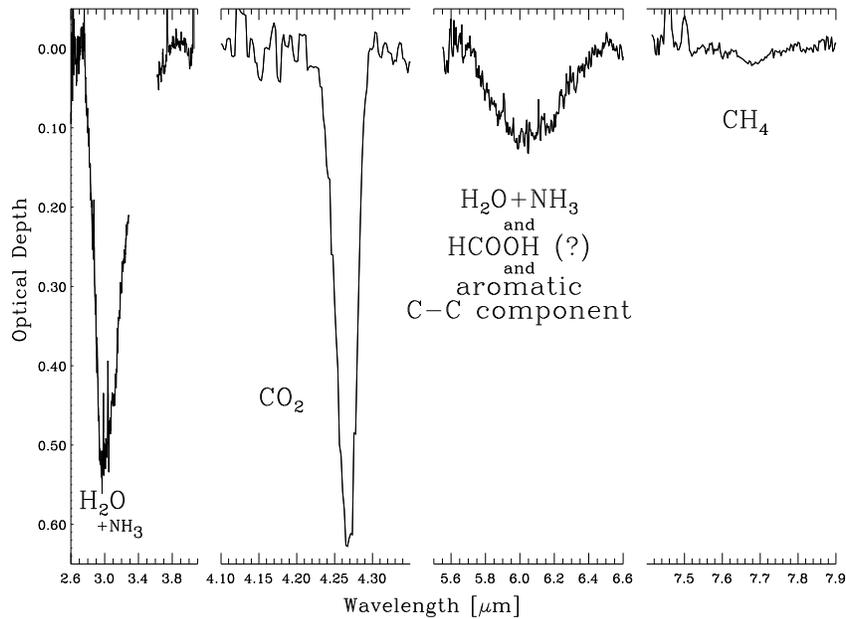,height=3.3in,angle=90}}
\figcaption{Molecular cloud dust absorption features identified in Sgr A*. 
In order to highlight the molecular cloud features, the 3.29--3.62 region has
been removed from the ``ice'' profile.  All data are AOT1, speed 4, except
for the 7 \micron\ region which is AOT6.
\label{fig:sgra_molecular}}
\end{figure}

Molecular cloud absorption features of solid-state \water, CO$_2$, and
\methane\ along the line of sight to the Galactic Center are assigned based on
their previous  identification in known molecular cloud sources.  Consistent
with the uneven distribution of molecular cloud material, these features vary
in depth across the GC field, whereas absorption features due to dust in the 
diffuse interstellar medium are observed at approximately constant depth
(\S\ref{aliphatics}).  It is evident from the depth of the 3.0 \micron\ ice
feature (Fig.~\ref{fig:all_full}) that there is more molecular cloud material
along the line of sight toward Sgr A* than towards GCS3 and 4.  Thus, we use
the \sgra\ spectra to characterize the molecular cloud dust that is present
along the line of sight to the Galactic Center.  Optical depth plots of the
molecular cloud features are presented in Fig.~\ref{fig:sgra_molecular}; column
densities are listed in Table~3.  We  discuss these features in detail below.

\subsection{\water-ice at 3.0 and 6.0 \micron}\label{ices} 

\subsubsection{The 3 \micron\ feature}\label{water} 
\begin{figure}
\centerline{\psfig{figure=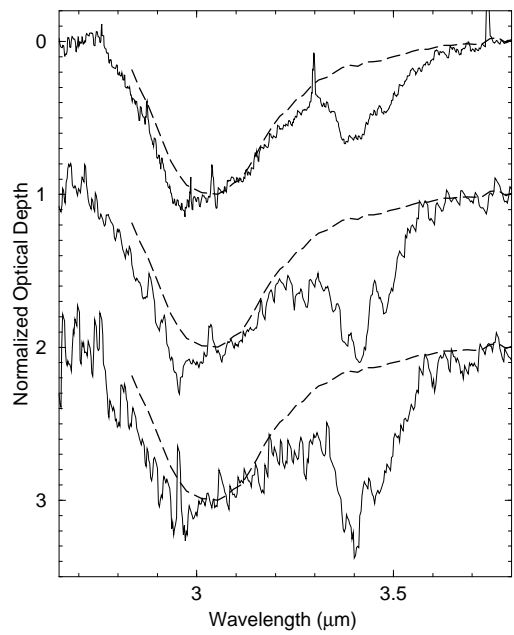,height=3.3in}}
\figcaption{Optical depth profiles of the 3 \micron\ ice feature for
Taurus dense cloud (Elias 16, dashed line; adapted from Smith \etal\ 1989)
compared with the Galactic Center sources \sgra\ (top), GCS3 (middle),
GCS4 (bottom).  All GC sources show excess blue absorption compared to the
Taurus ice profile.  Spectra have been normalized to have a depth of 1.0
at  3.05 \micron.
\label{fig:water}}
\end{figure}

The 3 \micron\ optical depth profiles for the GC sources
(Fig.~\ref{fig:water}) were determined by fitting a local second degree
polynomial in the 2.5--4.2 \micron\ region.   These profiles are very
similar to each other and at the same time distinctly different from the
\water-ice absorption features observed  towards molecular clouds
elsewhere in the Galaxy
\nocite{butchart_etal86,mcfadzean_etal89}(McFadzean \etal\ 1989; Butchart
\etal\ 1986).   Figure~\ref{fig:water} compares the 3 \micron\ water-ice
feature observed in the Taurus dark cloud in the line of sight towards the
background K-giant Elias 16 \nocite{smith_sellgren_tokunaga89} (from Smith
\etal\ 1989),  with those of the GC sources, SgrA*, GCS3 and 4.   Clearly,
the 3 \micron\ band in the Galactic Center shows a blue absorption wing,
peaking at about 2.95 \micron. (The aliphatic hydrocarbons absorption band
at 3.4 \micron\ is discussed in \S\ref{aliphatics}).  While initially this
difference in profile was taken to imply that this band might be due to
water of hydration in silicates rather than \water-ice 
\nocite{tielens_etal96,wada_sakata_tokunaga91,tielens_allamandola87b}
(Tielens \etal\ 1996; Wada \etal\ 1991; Tielens \& Allamandola 1987), the
absence of this band in local sightlines through the diffuse ISM sheds
some doubt on that interpretation \nocite{whittet_etal97}(e.g. Whittet
\etal\ 1997).  The discovery of solid CO$_2$ by ISO now firmly links the
3.0 \micron\ band in the GC to \water-ices \nocite{whittet_etal97}(de
Graauw \etal\  1996b; Lutz \etal\ 1996; Whittet \etal\ 1997; Gerakines
\etal\ 1999).  

We consider the possibility that the 3.0 \micron\ band
could be due to diluted mixtures of H$_2$O with other molecules,
such as CO, O$_2$, and \ammonia\ as suggested by
\nocite{schutte_greenberg89} Schutte \& Greenberg (1989).  Drawing upon
extensive laboratory studies by the Leiden
(http://www.strw.leidenuniv.nl/$\sim$lab/; Gerakines \etal\ 1995, 1996)
\nocite{gerakines_schutte_ehrenfreund96,gerakines_etal95} and
NASA/Ames Astrophysics Laboratory groups, we attribute the difference in
the 3.0 \micron\ profile between the GC and local molecular clouds to a
higher abundance of other species -- notably ammonia -- in the \water-ice
in the former relative to the latter regions.   Possibly, this increased
abundance of NH$_3$-ice in the GC region relative to \water-ice reflects
the increase in the N/O elemental abundance ratio with decreasing
galacto-centric radius (Simpson \etal\ 1995; Rubin \etal\ 1988).
\nocite{simpson_etal95,rubin_etal88} A detailed discussion of the 
laboratory results for \water:\ammonia\ mixtures will be discussed in a
subsequent paper (J.\ Chiar \& M.\ Bernstein, in preparation).   We describe the  laboratory
fits to the 3 \micron\ profile, and determination of \water-ice column 
density along with the discussion of the 6 \micron\ ice 
profile below. 

\subsubsection{The 6 \micron\ feature}\label{nightmare}
\begin{figure}
\centerline{\psfig{figure=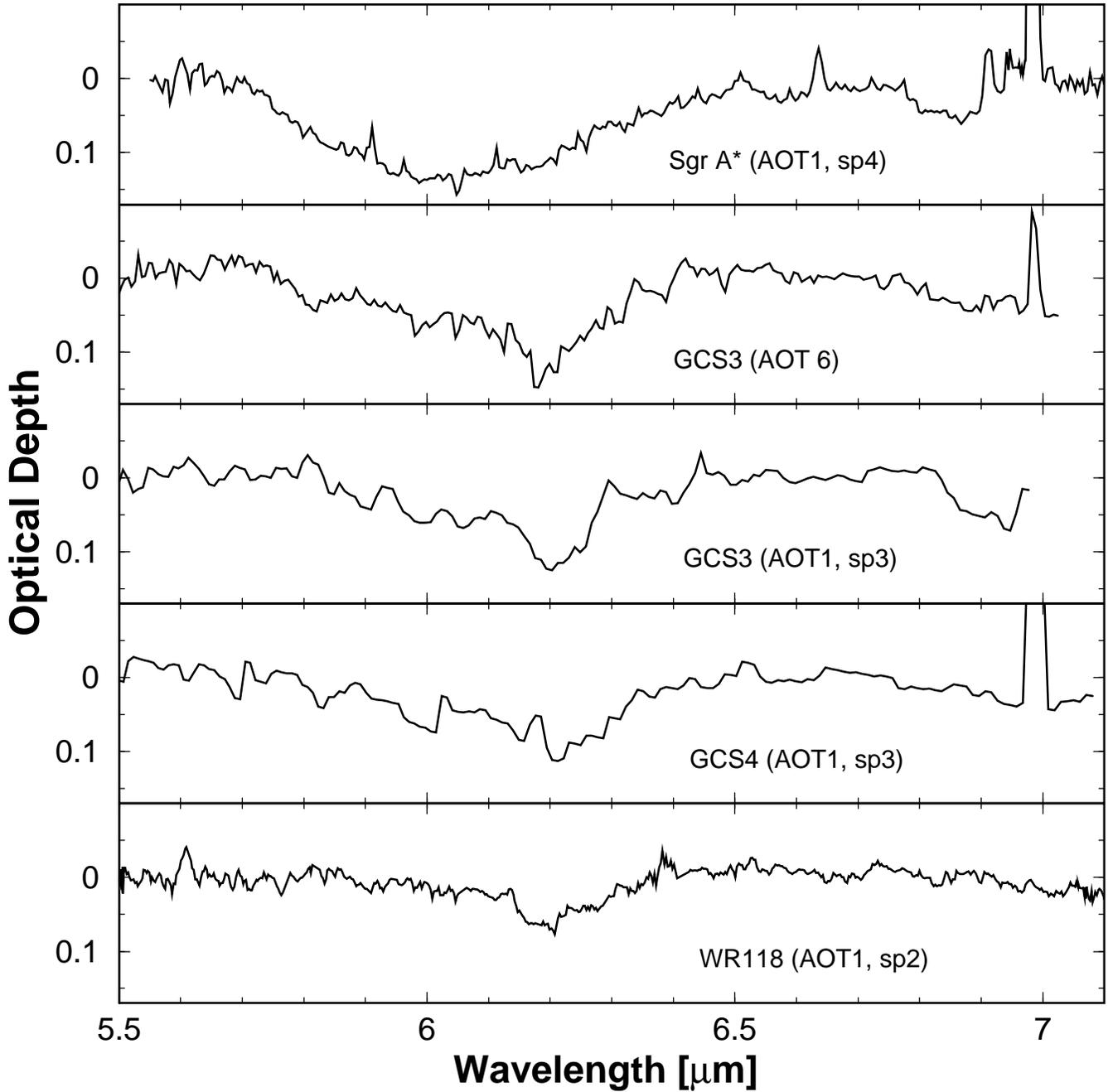,height=7in}}
\figcaption{The 6.0 \micron\ region of Sgr A*, GCS3, GCS4, and WR 118. 
The spectrum of WR118 is from Schutte \etal\ (1998).  Narrow absorption
lines (FWHM$\sim0.01$\micron) in the spectrum of GCS3 (and \sgra)
are due to cold gas-phase \water\ along the line of sight 
(Moneti \& Cernicharo 2000).
\label{fig:gcs3_6mic}}
\end{figure}

The 5.5--6.5 \micron\ spectral region contains the vibrational modes of
several molecules of astrophysical interest: O-H bend in \water\ (6.0
\micron), C=O stretch in carbonyl groups (occurring in aldehydes, ketones,
and carboxylic acids; 5.7--5.9 \micron), and C-C stretch in aromatic
materials (6.2 \micron). The spectra of all GC sources show an absorption
feature around 6.0 \micron\ (Fig.~\ref{fig:gcs3_6mic}).  Sources in or
behind local molecular clouds also show a feature at 6.0 \micron\
(Fig.~\ref{fig:ngc_sgra}) due
mostly to the O-H bending mode in \water-ice, although trace amounts of
other molecules may contribute to the absorption
\nocite{keane_etal00,schutte_etal96b}(Keane \etal\  2000; Schutte \etal\ 
1996b; Tielens \& Allamandola 1987).  In contrast, Galactic Wolf-Rayet
WC-type stars show an absorption feature at 6.2 \micron\ attributed to the
C-C stretch in aromatic materials (Schutte \etal\ 1998; see also
Fig.~\ref{fig:gcs3_6mic} and \S\ref{arom6}).  Visual inspection of
Fig.~\ref{fig:gcs3_6mic} reveals that the 6 \micron\ profile of Sgr A* is
strikingly different than those of GCS3 and 4, the latter two instead show
a similarity to the 6.2 \micron\ band observed towards late-type dusty WC
stars such as WR118  (cf.\ \S\ref{aromatics} and Schutte \etal\ 1998).
Given the greater depth of the 3.0 \micron\ ice band  in \sgra, compared
to GCS3 and GCS4 (Fig.~\ref{fig:all_full}), the 6.0 \micron\ band of this
source is likely to contain a dominant ice component.    

\begin{figure}
\centerline{\psfig{figure=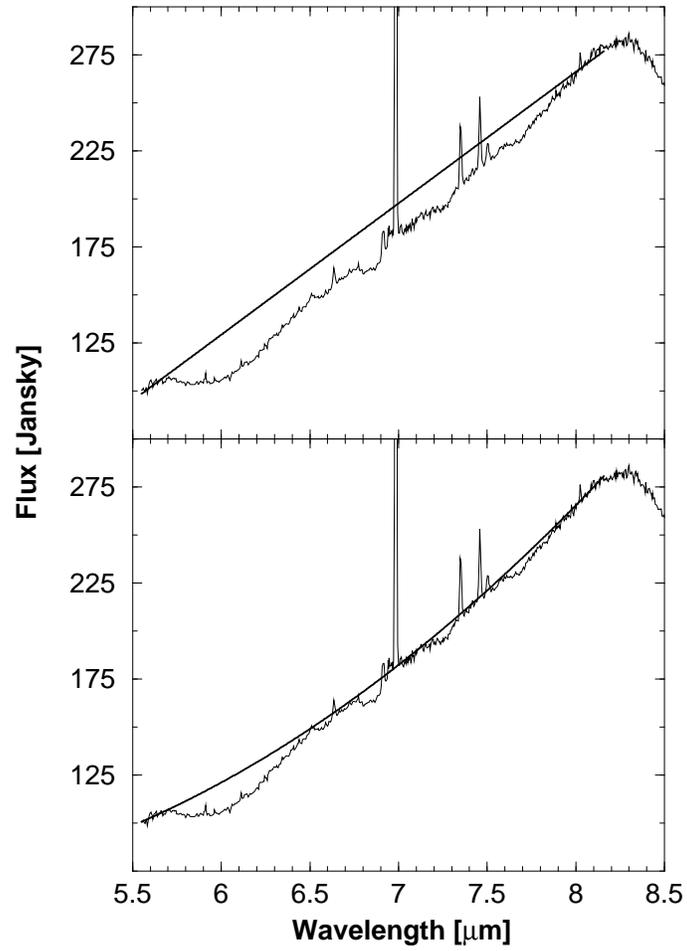,height=5in}}
\figcaption{5.5--8.5 \micron\ flux spectrum of \sgra\ showing second [top] and
first [bottom] degree continuum fits.  The resulting optical depth spectra
are shown in Figure~\ref{fig:ngc_sgra}.
\label{fig:sgra_cont}}
\end{figure}

\begin{figure}
\centerline{\psfig{figure=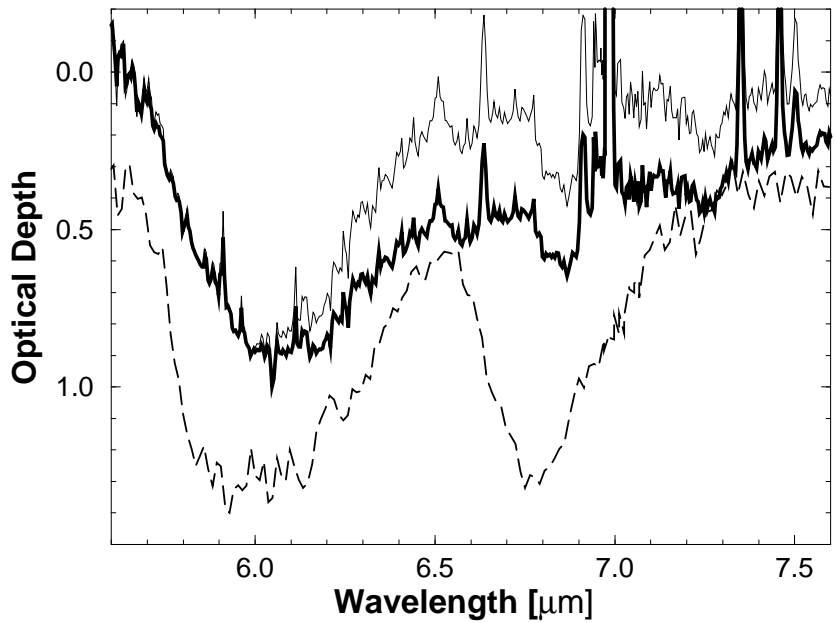,height=3.3in,angle=-90}}
\figcaption{Comparison of the 6 \micron\ feature in \sgra\ using two
different continuum fits as shown in Fig.~\ref{fig:sgra_cont}.  Optical
depth spectra resulting from the second degree polynomial fit [thin solid
line] and  first degree polynomial fit [thick solid line] are shown. The
optical depth spectrum of the protostellar object \ngc\ [dashed line;
Keane \etal\ 2000] is shown for comparison, and is offset for clarity. All
three spectra have been scaled to 1 at their peak optical depth. 
\label{fig:ngc_sgra}}
\end{figure}

Determination of the 6 \micron\ profile for \sgra\ depends heavily on the
choice of continuum, leading to uncertainties in the red side of the
profile and peak optical depth, and minor changes in the peak
position. We carried out first and second degree polynomial fits to the
5.5--8.5 \micron\ spectrum of \sgra\ in order to quantify the differences
in the profiles resulting from each.  Figs.~\ref{fig:sgra_cont} and
\ref{fig:ngc_sgra} show two possible continuum choices and the resulting
optical depth spectra, respectively.  Peak positions  of $\lambda\sim6.05$
(second degree polynomial fit) and 6.10 \micron\ (first degree polynomial
fit), and  peak optical depths (at 6.05 \micron) of 0.20 (first
degree) and 0.14 (second degree) are obtained. We adopt the second
degree continuum fit, as this provides a good estimate of the continuum
for the absorption features at 6.0, 6.85, and 7.25 \micron.\footnote{A
local continuum in the 7.3--8.0 \micron\ region is used for the \methane\
feature at 7.68 \micron. See \S\ref{methane} for further discussion.}  For
consistency, we also adopt second degree polynomial continuum fits for
GCS3 and 4; the optical depth spectra are shown in
Fig.~\ref{fig:gcs3_6mic}.

\begin{figure}
\centerline{\psfig{figure=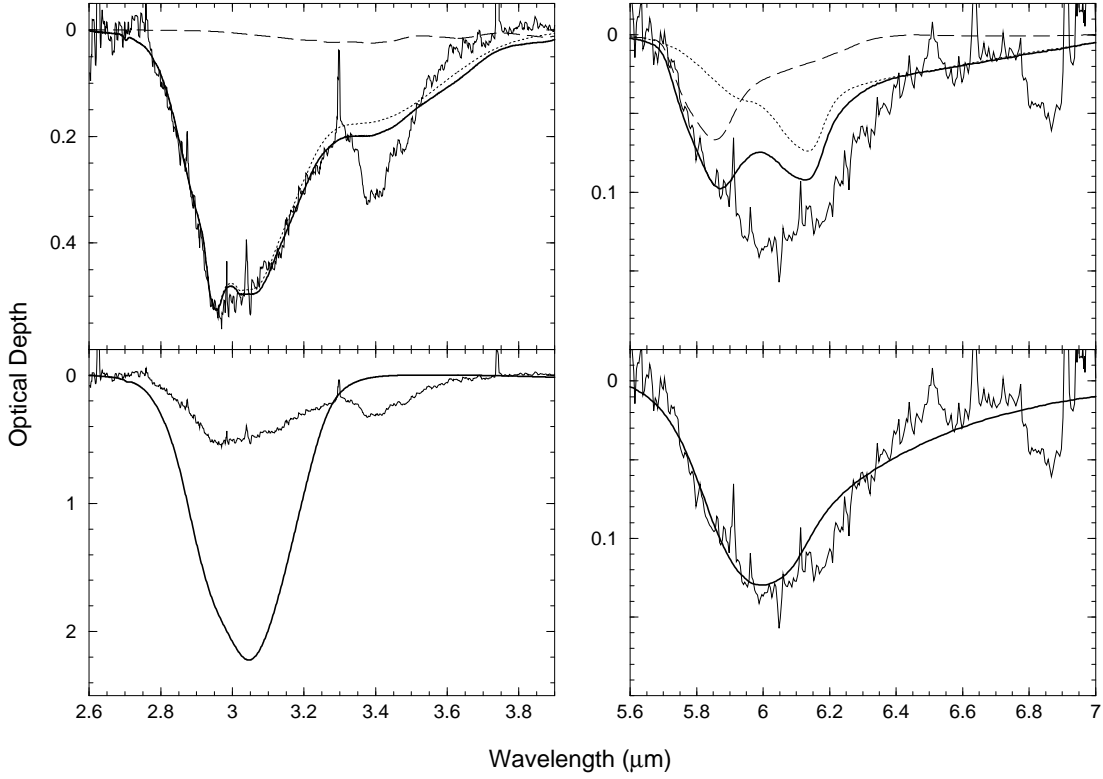,height=4in,angle=-90}}
\figcaption{The 2.6--3.9 \micron\ [left panels] and 5.6--7 \micron\ 
[right panels] optical depth spectra for  \sgra.  The top panels show the
result of the fitting procedure when both profiles are used to constrain
the fit.  The ``best'' fit is obtained with: H$_2$O:NH$_3$:CO$_2$ 
(3:1:0.03, 15K) (dotted line) and HCOOH at 10K (dashed line).  The sum of
the two laboratory mixtures is shown by the smooth solid line. The bottom
panels show the result of the fitting procedure when only the 6 \micron\
profile is considered.  In this case, although the 6 \micron\ band is
reasonably matched with pure \water\ at 30\,K (smooth solid line), the
depth of  the 3 \micron\ profile is overestimated by the laboratory
spectrum [bottom, left panel]
\label{fig:bigfits}}
\end{figure}

\begin{figure}
\centerline{\psfig{figure=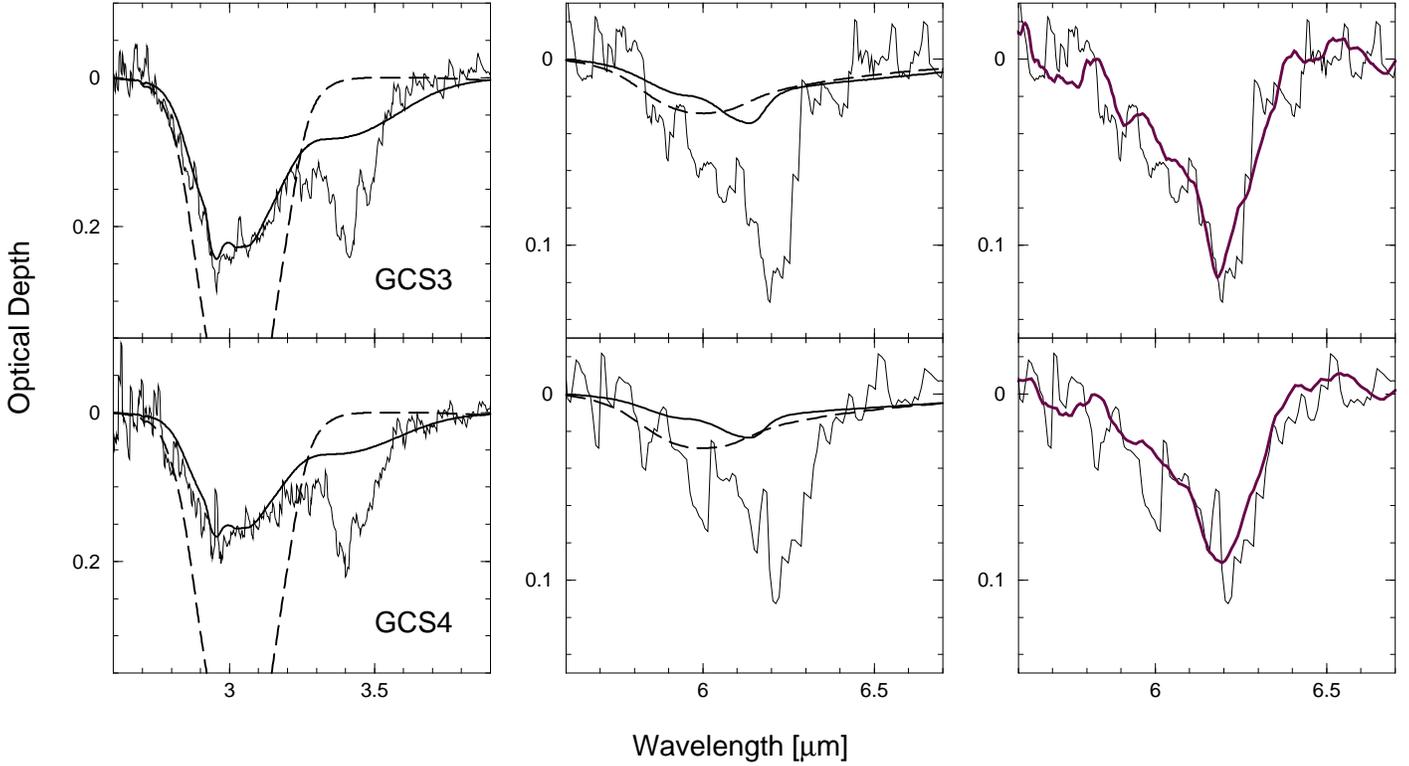,height=4in,angle=-90}}
\figcaption{ 
The 2.6--3.9 \micron\ [left panels] and 5.6--6.7 \micron\ [center and
right panels] spectral region of  GCS3 [top panels] and  GCS4 [bottom
panels]. Comparison of the laboratory mixtures  H$_2$O:NH$_3$:CO$_2$
(3:1:0.03, 15K) [smooth solid line] and pure \water\ at 30K [dashed line]
are shown for the 3 and 6 \micron\ spectral regions in the left and center
panels, respectively.  Similar to \sgra, the  H$_2$O:NH$_3$:CO$_2$ mixture
represents these profiles better than pure \water-ice.  The right panels
show the result of summing the H$_2$O:NH$_3$:CO$_2$ mixture with a
WR118-like 6.2 \micron\ component [heavy solid line].
\label{fig:quints}}
\end{figure}

Rather than identifying the carrier of the 6 \micron\ feature in the GC
sources in isolation, we  use the 3 \micron\ profile as a constraint for
the nature and abundances of the carriers of the corresponding 6 \micron\
feature.  Most of the 6 \micron\ absorption in protostellar objects is
explained by simple amorphous \water-ice mixtures, so we first attempt to
match the GC 3- and 6-\micron\ profiles with only pure \water-ice (using
laboratory data from Leiden Observatory's database at:
http://www.strw.leidenuniv.nl/$\sim$lab/).   This procedure shows  that
fitting  the 6 \micron\ band with pure \water-ice causes the optical depth
of the 3 \micron\ band to be overestimated by a factor of 2--3.5 for all 3
objects.  This is demonstrated in  Fig.~\ref{fig:bigfits} (bottom panels)
for SgrA* and in Fig.~\ref{fig:quints} for GCS3 (top panels) and  4
(bottom panels).  The mixture used, amorphous \water\ at 30\,K is a
realistic analog for ices in these sightlines,  and provides the best fit
to the \sgra\ 6 \micron\ profile when the 3 \micron\ spectrum is
ignored.   Such a discrepancy between the 3 and 6 \micron\ \water-ice
optical depth has been noted for some massive young stellar objects (Gibb
\etal\ 2000; Keane \etal\ 2000).  For these sources scattering at near-IR
wavelengths in the disk-like environment of the protostar may reduce the
apparent optical depth at 3.0 \micron\  \nocite{pendleton_tielens_werner90}
(Pendleton \etal\ 1990).  For the Galactic Center sources, ice absorption
is not local to the sources but rather occurs along the line-of-sight and,
hence, this effect should not play a role.  Thus, we conclude that,
despite the good fit, the 6.0 \micron\ band cannot be due to \water-ice
alone.

The 3 \micron\ band towards \sgra\ shows an excess blue absorption wing
relative to other observed interstellar ice bands (cf.
Fig.~\ref{fig:water}).  Likely, this is due to the presence of \ammonia\
in the ice (e.g., Schutte \& Greenberg 1989). There is evidence that the 6
\micron\ profile in protostellar objects is also due to a mixture of other
molecules with \water\ \nocite{gibb_etal00}(Schutte \etal\ 1996b; Keane
\etal\ 2000; Gibb \etal\ 2000).   Thus, we favor a multicomponent ``fit''
to the 3- and 6-\micron\ ice features, and have used the extensive
databases at Leiden Observatory and NASA-Ames Astrophysical Laboratories
to determine the carriers (in addition to \water-ice) responsible for the
two features.  These databases contain mixtures of \water-ice with
\methanol, \methane, O$_2$, \formald, HCOOH, CO, and CO$_2$ as minor
components, as well as pure ices of these molecules, at temperatures
ranging from 10--245\,K (http://www.strw.leidenuniv.nl/$\sim$lab/;
J.\ Chiar \& M.\ Bernstein, in preparation).  A formal chi-squared fitting
routine is used to fit the profiles in the SgrA* spectrum.   The ``best''
fit, shown in Fig.~\ref{fig:bigfits} (top panels), consists of
\water:\ammonia:CO$_2$ (100:30:6, 15\,K) and HCOOH at 10K.  Summing these
mixtures provides an impressive fit to the 3 \micron\ feature including
the so-called long-wavelength wing (3.4 \micron\ absorption is due to
diffuse ISM dust; this region is excluded from the fit), and can account
for much of the absorption at 6 \micron.   Formaldehyde (H$_2$CO) has been
discussed as a candidate for the red side of the 6 \micron\ feature in
protostellar objects \nocite{schutte_etal96a}(Schutte \etal\ 1996a; Keane
\etal\ 2000); the \sgra\ profile (which is tightly constrained on the red
side, regardless of continuum choice), limits the amount of \formald\
relative to \water-ice to $\sim2\%$.  

\begin{figure}
\centerline{\psfig{figure=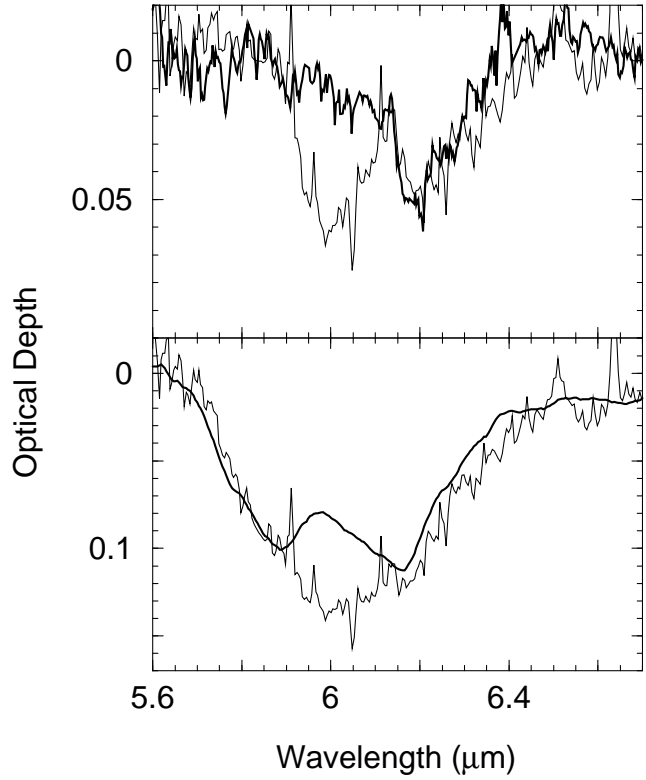,height=4in}}
\figcaption{[Top panel] The 6 \micron\ residual for \sgra\ obtained after
subtracting the best-fitting laboratory mixtures shown in
Fig.~\ref{fig:bigfits}.  The heavy line represents the WR118 6.2 \micron\
optical depth spectrum, divided a factor of 1.25.  The bottom panel shows the
``fit'' resulting from the summation of the best-fitting laboratory ices
and the scaled WR118 6.2 \micron\ feature.
\label{fig:sgra_residual}}
\end{figure}

Residual absorption resulting from subtracting the ``fit'' from the
astronomical data is shown in Fig.~\ref{fig:sgra_residual} (top panel). 
Much of this remaining absorption is accounted for by a component at 6.2
\micron, similar to the WR118 6.2 \micron\ profile
(Figs.~\ref{fig:gcs3_6mic}, \ref{fig:sgra_residual}). See \S\ref{arom6}
for a detailed discussion of this component.  Only a small portion of the
6 \micron\ profile remains unaccounted for in the \sgra\ spectrum
(Fig.~\ref{fig:sgra_residual}, bottom panel). Additional  organic acid
mixtures, such as HCOOH at a range of temperatures and host matrices, are
needed to resolve this issue.

The weakness of the  6 \micron\ ice components in GCS3 and 4 make a
chi-squared analysis of the 3- and 6-\micron\ bands unwarranted (see the
middle panels in Fig.~\ref{fig:quints}), however  their 3-\micron\
profiles and the relative depths of their 3- and 6-\micron\ bands can be
used to help constrain the nature of the ices.   Pure \water\ does not
provide a satisfactory match to the 3- and 6-\micron\ profiles, as
discussed above (Fig.~\ref{fig:quints}).  The 3 \micron\ profiles of these
sources are well-matched by the same \water:\ammonia:CO$_2$ mixture used
to fit the \sgra\ profiles, suggestive of a consistent \ammonia/\water\
ratio for dense clouds towards the Galactic Center.  A 6.2 \micron\
WR118-like component accounts for the remainder of the 6 \micron\ profile,
as for \sgra.  Fig.~\ref{fig:quints} shows that all of the 6 \micron\
absorption in  the Quintuplet spectra is accounted for by the sum of the
ice mixture (\water:\ammonia:CO$_2$, 15\,K) and a 6.2 \micron\ ``diffuse
ISM'' component (represented by the WR118 spectrum).  Detailed discussion
of the 6.2 \micron\ feature is deferred to \S\ref{arom6}.  

To summarize, using the available databases of interstellar ice analogs,
the 3- and 6-\micron\ profiles observed toward \sgra, GCS3, and GCS4 are
best represented  by the ice mixture, \water:\ammonia:CO$_2$ (100:30:6,
15K), with a small amount of HCOOH at 10K (for \sgra), in addition to a
diffuse ISM component at 6.2 \micron.  Pure \water\ cannot account for the
shape of the 3 \micron\ profile or the relative depths of the 3 and 6
\micron\ features.  Water-ice column densities (Table~3) are determined by
the fit to the 3 \micron\ profile (Fig.~\ref{fig:bigfits} [top left,
dotted line], 
Fig.~\ref{fig:quints} [left panels, solid line]), using $N =
\int\tau_{\nu}d\nu/A$, and \nocite{hagen_tielens_greenberg81}
$A=2\times10^{-16}$ cm molecule$^{-1}$ (Hagen \etal\ 1981).  Column
density for HCOOH is determined using the 6 \micron\ fit
(Fig.~\ref{fig:bigfits} [top right, dashed line]), and
$A=6.7\times10^{-17}$ cm molecule$^{-1}$ \nocite{marechal87} 
(Mar\'echal 1987).

\subsection{CO$_2$ and \methanol-ice, and CO}\label{otherices}
Solid CO$_2$ absorption features at 4.27 \micron\ (stretching mode) and 15.2
\micron\ (bending mode)  have been discussed extensively by Gerakines \etal\
(1999).  Fits to the CO$_2$ profiles showed that toward the sources in the
present paper, CO$_2$ ice is embedded in a mainly polar matrix at temperatures
$<50$\,K.

We searched for solid \methanol\ absorption features at 3.54 \micron\ (C-H
stretch), 3.9 \micron\ (combination mode), and 6.85 \micron\ (C-H
deformation mode) in the \sgra\ spectrum. The C-H stretch mode occurs in
the red wing of the 3.4 \micron\ aliphatic hydrocarbon feature.  In order
to obtain a limit on the depth of a possible \methanol\ feature at 3.54
\micron, a local second degree polynomial was fitted between 3.4--3.7
\micron.  A second degree polynomial continuum was fitted in the 3.6--4.1
\micron\ region to estimate  the 3.9 \micron\ feature depth.  Although we
detect a feature centered at 6.85 \micron\ towards SgrA*, its width is too
narrow to be due to \methanol\ (see \S\ref{aliphatics}), thus the
continuum on either side of the observed 6.85 \micron\ feature is used to
deduce a limit for the \methanol\ feature.  Limits on the optical depths
are $\tau_{3.54} < 0.01$,  $\tau_{3.9} < 0.001$, $\tau_{6.85} < 0.005$. 
Using the integrated absorbance values for pure \methanol\
($A=5.3\times10^{-18}, 2.8\times10^{-18}, 1.2\times10^{-17}$ cm\,
molecule$^{-1}$, for the 3.54, 3.9, and 6.85 \micron\ features,
respectively; Hudgins \etal\ 1993), the average calculated  \methanol\
column density is $<0.5\times10^{17}$ cm$^{-2}$, resulting in an abundance
of $<4\%$ relative to \water-ice.  This is similar to the value,
$N$(\methanol)/$N$(\water) $<5\%$, found for the quiescent Taurus dark
cloud \nocite{chiar_adamson_whittet96} (Chiar, Adamson, \& Whittet 1996),
and significantly less than the abundance found for some high mass
protostars where $N$(\methanol)/$N$(\water) is as high as  30\%
\nocite{allamandola_etal92,dartois_etal99, skinner_etal92} (Allamandola
\etal\ 1992; Skinner \etal\ 1992; Dartois \etal\ 1999).

Carbon monoxide is present primarily in the gas-phase in these lines of
sight.  Limiting values for the solid CO column densities derived in this
work are listed in Table~3.  These values  agree with those independently
analyzed by Gerakines \etal\ (1999).

\subsection{\methane-ice}\label{methane}
\begin{figure}
\centerline{\psfig{figure=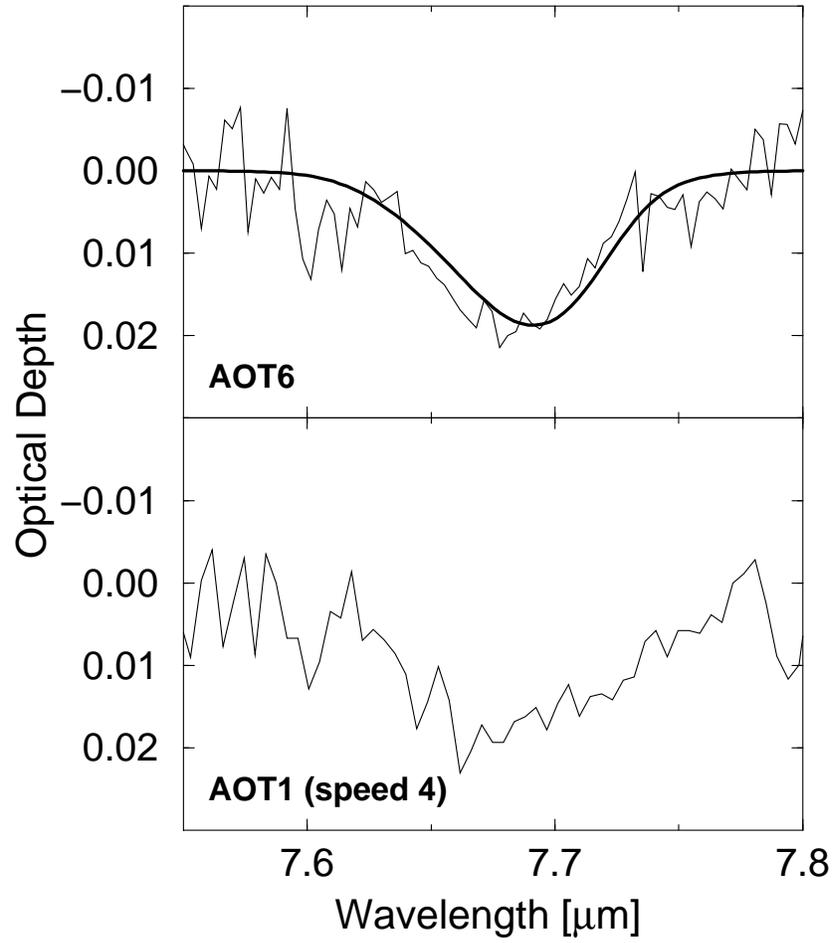,height=5in}}
\figcaption{The 7.55--7.8 \micron\ optical depth spectrum of  Sgr A* shown
with best-fitting laboratory mixture: \water:\methane\ (3:1, 10K) (smooth
solid line). 
\label{fig:methane}}
\end{figure}

The deformation mode of solid methane (\methane) occurs near 7.68 \micron,
and has been detected in three protostellar sources to date
\nocite{boogert_etal96} (d'Hendecourt \etal\ 1996; Boogert \etal\ 1996,
1998). We detect this absorption feature in the \sgra\ and \sgraw\ spectra
with $\tau_{7.68}\simeq0.02$ (Fig.~\ref{fig:methane}).   Fits to the
higher quality \sgraw\ AOT6 spectrum were carried out using the database
at Leiden Observatory \nocite{boogert_etal97}(Boogert \etal\ 1997).  This
spectrum is best-fitted by \methane\ embedded in a polar matrix containing
molecules such as \water\ and/or \methanol\ at temperatures ranging from
10--80\,K.  Nonpolar mixtures do not provide a good fit as they peak at
shorter wavelengths and have more narrow
profiles than the observed feature.  Our result is consistent with recent  studies of solid
\methane\ toward protostellar objects which have shown that \methane\ 
embedded in a polar matrix provides the best fit to the observed profile 
(Boogert \etal\ 1996; d'Hendecourt \etal\ 1996).  We adopt the
matrix-independent integrated absorption strength, $A=7.3\times10^{-18}$
cm molecule$^{-1}$ to compute the column density of \methane\ towards
SgrA*: $N$(CH$_4)=3\pm1\times10^{16}$ cm$^{-2}$, resulting in
$N$(CH$_4$)/$N$(H$_2$O) = $0.02\pm0.01$, within the range, 0.4-4\%, found
for protostellar objects  
\nocite{boogert_etal96,boogert_etal98,dhendecourt_etal96}(Boogert \etal\
1996; d'Hendecourt \etal\ 1996).

The peak optical depth of the C-H stretching mode of solid  \methane,  at
3.32 \micron, is expected to be the same as that of the deformation mode
\nocite{hudgins_etal93}(Hudgins \etal\ 1993; Boogert \etal\ 1997).  Since
this feature falls  in the blue wing of the 3.4 \micron\ aliphatic
hydrocarbon feature (\S\ref{aliphatics}),  its presence cannot be
confirmed.

Assuming a similar $N$(CH$_4$)/$N$(H$_2$O) ratio for GCS3 and GCS4 as for
\sgra, a feature with optical depth $\tau_{7.68}<0.01$  is expected
for the C-H deformation mode.  The current data do not have sufficient S/N
in this wavelength region to detect such a weak feature; limiting values
for $\tau_{7.68}$ and $N$(\methane) are given in Tables~2 and 3,
respectively.

\section{The Diffuse ISM features}\label{diffuse}

Aliphatic hydrocarbons are known to be a widespread component of the
diffuse ISM as shown by the ubiquitous presence of the 3.4 \micron\ C-H
stretch absorption feature in all studied sightlines.  ISO-SWS data have
now given us irrefutable evidence for the corresponding deformation modes
at longer wavelengths.  Contrarily, aromatic hydrocarbons, which have C-H
and C-C stretch modes at $\sim$3.3 and 6.2 \micron, respectively, are not
yet well-established as a universal component of the diffuse ISM.  We
discuss these hydrocarbon features toward the GC in the following sections.
Aliphatic hydrocarbon features detected toward \sgra\ are shown in
Fig.~\ref{fig:sgra_diffuse}; Fig.~\ref{fig:gcs3_organics} displays both
aliphatic and aromatic hydrocarbon absorption features observed
toward GCS3.

\subsection{The 3.4, 6.85, and 7.25 \micron\ aliphatic hydrocarbon
features}\label{aliphatics}

\begin{figure}
\centerline{\psfig{figure=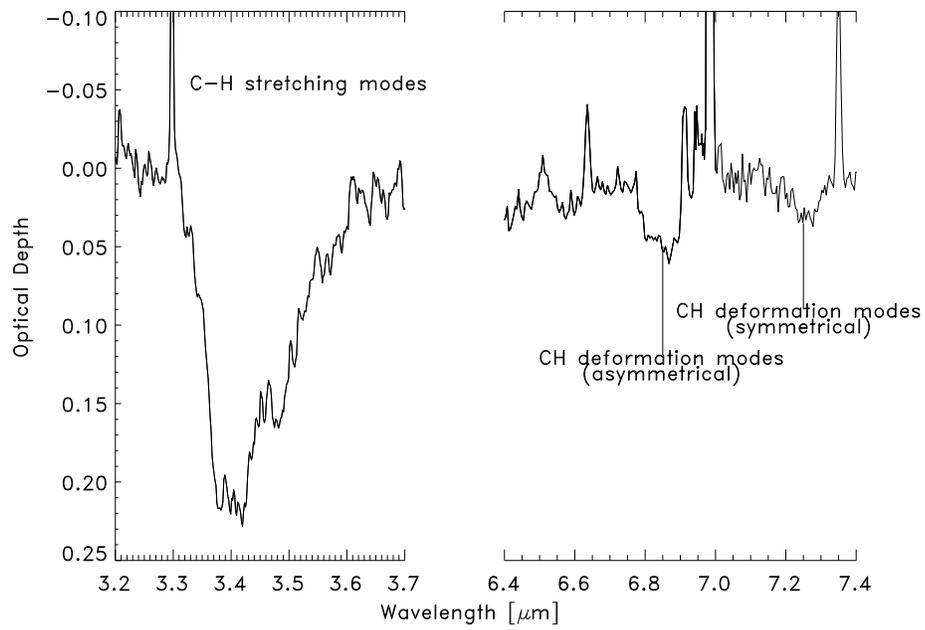,height=3.5in,angle=90}}
\figcaption{Aliphatic hydrocarbon absorption features identified in Sgr A* (AOT1
data).  Shown are the well-known 3.4 \micron\ C-H stretch
absorption feature and its corresponding deformation modes at 6.85
and 7.25 \micron.
\label{fig:sgra_diffuse}}
\end{figure}

\begin{figure}
\centerline{\psfig{figure=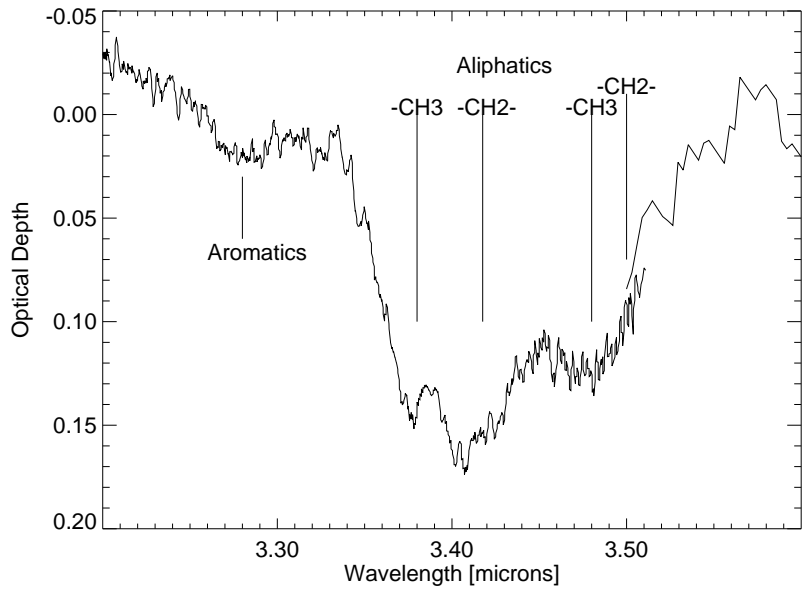,height=3.3in,angle=90}}
\figcaption{Hydrocarbon features toward GCS3, combining AOT6 (3.2-3.5
\micron) and AOT1 (3.5-3.6 \micron) data.  The optical depth spectrum
was created by fitting a local second degree polynomial continuum to the
combined spectra in 3.1--3.7 \micron\ region.
Aliphatic hydrocarbon
subfeatures at $\sim$ 3.38 \micron\ (-CH$_3$ asymmetric stretch),
3.42 \micron\ (-CH$_2$- asymmetric stretch), 3.48 \micron\ (-CH$_3$
symmetric stretch), and 3.50 \micron\ (-CH$_2$- symmetric stretch) 
are indicated.  The latter two subfeatures are blended.
Also present is an aromatic hydrocarbon feature at 3.28 \micron.     
\label{fig:gcs3_organics}}
\end{figure}

Saturated aliphatic hydrocarbons exhibit CH stretching modes at $\sim3.4$
\micron\ (methyl and methylene groups), and CH deformation modes at
$\sim6.9$ \micron\ (methyl and methylene groups) and $\sim7.3$ \micron\
(methyl group).  The profile of the 3.4 \micron\ absorption features
toward Sgr A* (Fig.~\ref{fig:sgra_diffuse}), GCS 3
(Fig.~\ref{fig:gcs3_organics}), and 4 is consistent with previous
ground-based observations: substructure in the feature is identified with
the CH stretch in methyl (-CH$_3$) and methylene (-CH$_2$-) groups, with
an average ratio of \methylene/\methyl$\simeq2-2.5$, in saturated
aliphatic hydrocarbons (Pendleton \etal\ 1994; Sandford \etal\ 1991). 
Subfeatures due to the symmetric and asymmetric vibrational stretching
modes of the CH groups are indicated in  the 3.2--3.9 \micron\ 
optical depth spectrum for GCS3 (Fig.~\ref{fig:gcs3_organics}). 

The \sgra\ spectrum shows clear absorption features at 6.85 \micron\ 
and 7.25 \micron\  (Fig.~\ref{fig:sgra_diffuse}) due to the asymmetric and
symmetric CH deformation modes, respectively. The 6.85 \micron\ band is
also weakly present in the spectrum of GCS3, and may be present in the
spectrum of GCS4, consistent with the relatively weaker 3.4 \micron\
absorption features (Table~2, Fig.\ref{fig:gcs3_6mic}).  We do not attempt
to quantify the central wavelengths of the weak 6.85 \micron\ features in
the GCS3 and 4 spectra; within reasonable uncertainties they are similar
to that in \sgra.  The signal-to-noise of  the GCS3 and 4 spectra in the 7
\micron\ region is not sufficient to detect the 7.25 \micron\ band; limits
are given in Table~2.   Using the \sgra\ spectrum, and assuming
-\methylene-/-\methyl\ = 2.5, we calculated the expected ratio of
integrated intensities for the 3.4, 6.85 and 7.25 \micron\   bands for
branched and normal saturated aliphatic hydrocarbons from values by Wexler
(1967) (Table~4).    The 6.85 and 7.25 \micron\ feature profiles are
estimated  using Lorentzian curves with FWHM = 0.12 \micron\ (26
cm$^{-1}$), and 0.10 \micron\ (19 cm$^{-1}$), respectively, assuming that
the 6.85 \micron\ feature is  partially filled in by emission features
(due to H$_2$ and [ArII]; see Lutz \etal\ 1996). Integrated intensities
for the astronomical data (Sgr A*) are determined using  the equation 
$\tau_{int} = \int\tau_{\nu}d\nu$, and are listed in Table~4.

\begin{figure}
\centerline{\psfig{figure=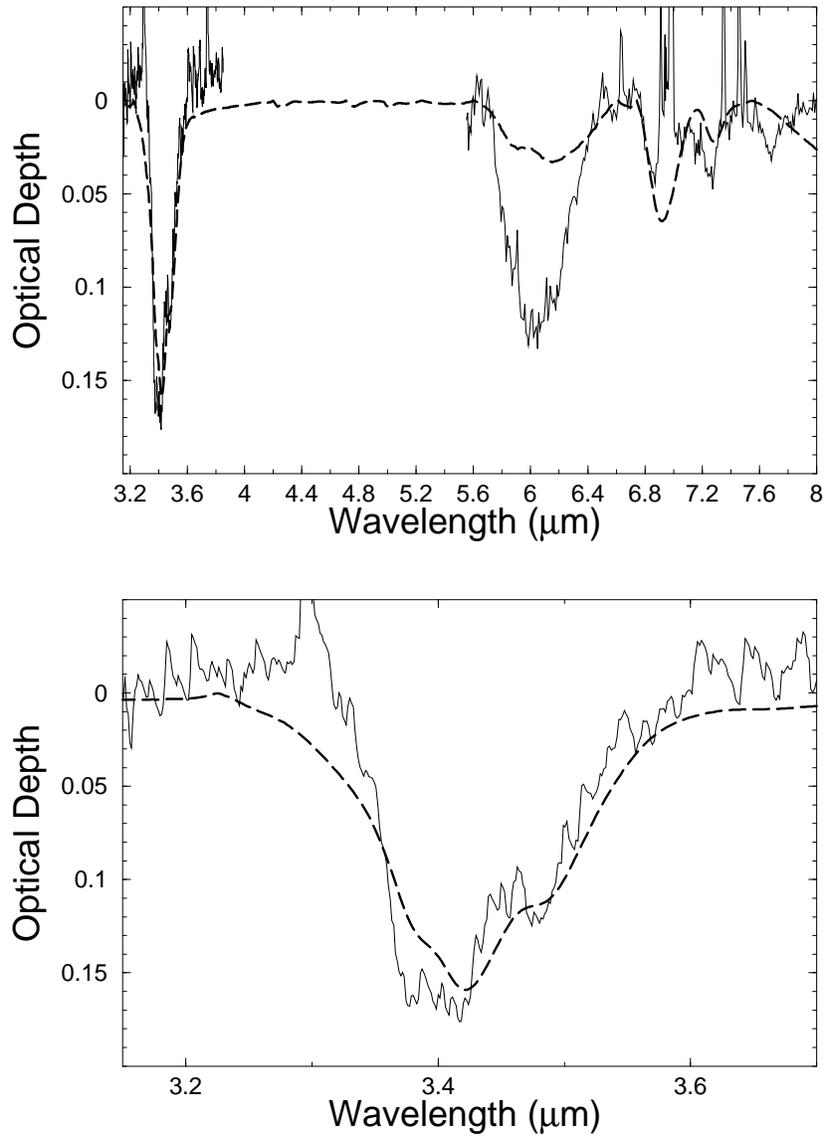,height=6in}}
\figcaption{Mid-IR spectrum of SgrA* compared with laboratory HAC (dashed
line; from Furton \etal\ 1999.  Bottom panel shows a close-up of the 3.4
\micron\ feature compared with HAC).
\label{fig:hac}}
\end{figure}

Integrated intensity ratios predicted for (branched) saturated aliphatic 
hydrocarbons are well-matched by those calculated from our \sgra\ spectrum
(Table~4).  Hydrogenated amorphous carbon (HAC) material can contain a
substantial fraction of saturated aliphatic hydrocarbon groups, and in
fact a HAC analog with   H/C = 0.5 and $sp^3/sp^2$ = 0.5  provides a
reasonable match to the relative strength of the 3.4, 6.85, and 7.25
\micron\ features in the Sgr A* spectrum
\nocite{furton_etal99}(Fig.~\ref{fig:hac}; Furton \etal\ 1999).   

Protostars also show a prominent feature at 6.85 \micron\ which appears to
be associated  with carriers residing in ices along the line of sight
(Keane \etal\ 2000).  The 6.85 \micron\ ice band, however, extends from
approximately 6.5--7.2 \micron, much broader that that observed towards
\sgra, GCS3 and 4 (Fig.~\ref{fig:ngc_sgra}).  The ice profile varies in
peak wavelength, width and depth for different protostellar objects, and
also correlates with the 3 \micron\ ice feature (Keane \etal\ 2000),
whereas no such strong variation in the feature is observed toward the GC,
even though there are clearly different amounts of molecular cloud ices.
Moreover, the 6.85 \micron\ optical depth is very similar in the three
Galactic Center sources despite the large differences in ice optical
depth.  Therefore, we attribute the 6.85 \micron\ feature to dust in the
diffuse interstellar medium, related to the 3.4 \micron\ aliphatic
hydrocarbon feature, rather than molecular cloud ices.

\subsection{Aromatic hydrocarbon features}\label{aromatics}

\subsubsection{The 6.2 \micron\ band}\label{arom6}
All the GC sources show evidence for the presence of a 6.2 \micron\ 
(FWHM $=34\pm5$ cm$^{-1}$) feature.  This feature has been
previously detected in ISO spectra of GC sources and 4 Wolf-Rayet stars
and is attributed to the C-C stretch in aromatic hydrocarbons (Schutte
\etal\ 1998).  

The 6.2 \micron\ profiles presented here were determined by first finding
a satisfactory fit to the 3- and 6-\micron\ ice features as described in
\S\ref{ices}, then subtracting this ``fit'' from the 6 \micron\ absorption
feature to obtain a residual spectrum (Figs.~\ref{fig:bigfits}, 
\ref{fig:quints}, \ref{fig:sgra_residual}).  The 6.2 \micron\ profile for
WR118 is used as a template ``diffuse ISM'' spectrum since this line of
sight is known to contain no molecular cloud material, based on
non-detections of typical molecular cloud features \nocite{vanderhucht_etal96}
(\water-ice and CO$_2$: van der Hucht \etal\ 1996; Schutte
\etal\ 1998). In the cases of GCS3 and GCS4, this procedure leads to a
residual which closely matches the (scaled) WR118 6.2 \micron\ profile.
Fig.~\ref{fig:quints} (right panels) shows the result of adding the scaled
WR118 profile to the ice fits for these sources, and shows that the 3 and
6 \micron\ profiles can be fully accounted for by ices plus a diffuse ISM
component.  In the case of \sgra, only the red side of the residual is
accounted for by a diffuse ISM (WR118-like) component.  The excellent
match of the red wing of the profiles gives us confidence that \sgra\ does
contain the same 6.2 \micron\ feature seen in GCS3, 4 and the WR stars.  

The 6.2 \micron\ feature is most likely carried by dust which is
unassociated with molecular cloud material  since it is detected in
lines-of-sight which are known to contain no ices (such as \water, CO$_2$;
Schutte \etal\ 1998).  In the Galactic Center, the source with the
strongest ice bands (\sgra) has the smallest 6.2 \micron\ feature. For all
the sources studied, there appears to be a trend of increasing 6.2
\micron\ feature depth with \av\ and $\tau_{9.7}$ (Schutte \etal\ 1998),
however the depth of the 6.2 \micron\ feature in the \sgra\ spectrum seems
to be anomalously weak with respect to its 9.7 \micron\ silicate optical
depth and its extinction, compared to sources in the solar neighborhood
(Schutte \etal\ 1998) and GCS3 and 4.  A depth of $\tau_{6.2}\sim0.13$
would be expected for \sgra\ if this quantity  was correlated with the
silicate depth.  In contrast, the observed depth is only 0.05 (Table~2).
Hence it seems that the 6.2 \micron\ feature is carried by an independent
component whose abundance varies with respect to that of the silicates. 
We note that the total visual extinction (\av) also seems to vary
independently of the 9.7 \micron\ silicate band \nocite{roche_aitken85}
(Roche \& Aitken 1985).

Presence of aromatic hydrocarbons can also be determined from observations
of the complimentary 3.3 \micron\ C-H stretch feature.   This feature is
discussed in more detail below.

\subsubsection{The 3.28 \micron\ aromatic hydrocarbon feature}\label{arom3}

The AOT6 spectrum of GCS 3 exhibits a weak absorption feature centered at
3.28 \micron, with $\Delta\nu=25\pm5$\,cm$^{-1}$, which we attribute to the
C-H stretch in aromatic hydrocarbons (Fig.~\ref{fig:gcs3_organics}). The
C-H stretch in solid methane occurs near 3.32 \micron, far from the
observed feature, therefore we can rule out methane as a possible
candidate for the absorption.  The lower resolution/signal-to-noise AOT1 
spectrum  is also consistent with the presence of absorption centered at
3.28 \micron.  Presence of this feature in the AOT1 spectrum of GCS4
cannot be ruled out.  An absorption feature centered at
$\lambda_0\sim3.25$ \micron\ ($\Delta\nu$=74\,cm$^{-1}$) has been detected
in molecular cloud sources and is attributed to aromatic hydrocarbon
molecules at low temperature (Sellgren \etal\ 1995; Brooke \etal\ 1999). 
The difference in central wavelength and width between these different
environments cannot be accounted for by uncertainties in the continuum
fit, but rather demonstrates the difference in the temperature and/or
carrier of the aromatic material in these regions. Although there is a
strong Pfund $\delta$ atomic hydrogen line at 3.30 \micron,  there is no
evidence for an underlying absorption feature at 3.28 \micron, to a limit
of $\tau_{3.28}<0.02$, in the SgrA* spectrum.   This limit is consistent 
with the weakness of the 6.2 \micron\ aromatic band in \sgra, thus we
conclude that the 3.28 \micron\ and 6.2 \micron\ feature
are due to the same material.

\begin{figure}
\centerline{\psfig{figure=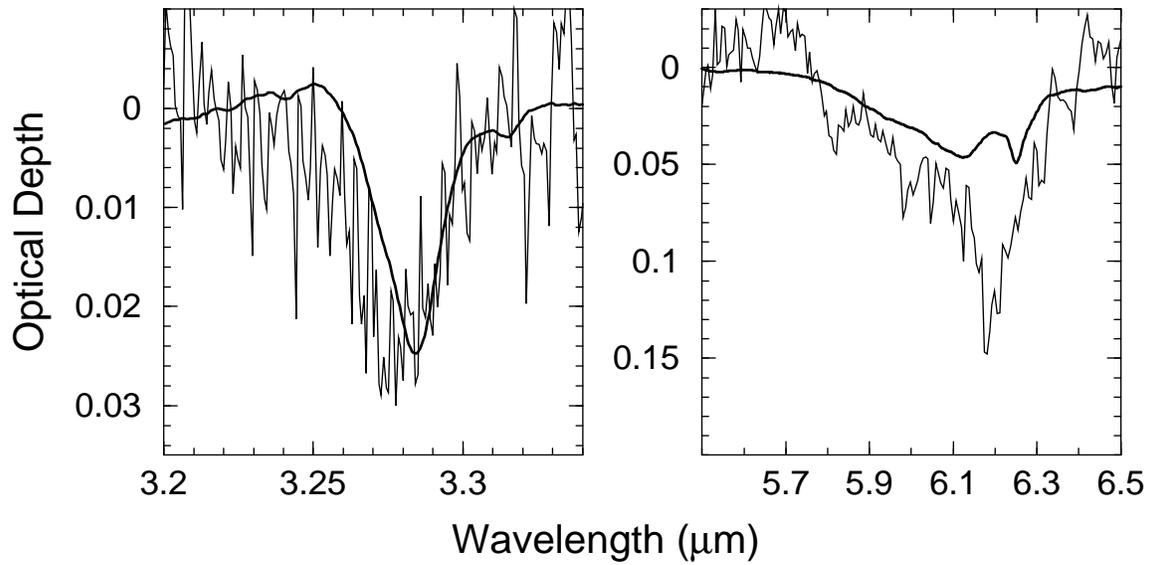,width=6in,angle=-90}}
\figcaption{The aromatic C-H stretch [left] and C-C stretch [right]
spectral regions in  GCS3 compared with a \water:C$_{16}$H$_{10}$ (14\,K)
laboratory mixture (smooth solid line).  The   6.2 \micron\ band in the laboratory mixture is
scaled based on the strength of the 3.28 \micron\ band.  It is apparent
that a PAH in ice mixture cannot account for most of the observed 6.2
\micron\ absorption. Spectra shown are AOT6.
\label{fig:sgra_pah}}
\end{figure}

Gas-phase PAHs are not a likely carrier of the 3.28 \micron\ absorption
feature: based on the temperature
dependence of the gas-phase PAH profile determined in the laboratory by
\nocite{joblin_etal95} Joblin \etal\ (1995), a temperature of 600-1300 K
is implied for the feature observed toward GCS3.  While such a high
temperature is consistent with PAH emission - indeed 3.28 \micron\ is also
the position of the unidentified infrared emission feature - to get a PAH
absorption feature would require an extraordinarily large column of hot
gas.   Comparison of the peak positions  of matrix-isolated laboratory
PAHs given by \nocite{hudgins_sandford98a,hudgins_sandford98b} Hudgins \&
Sandford (1998a,1998b) with our astronomical data show that neutral PAHs
such as pyrene frozen in ices may be a good candidate for the observed
feature. Laboratory studies carried out in the Astrophysics laboratory at
NASA Ames of pyrene in \water\ ice show a fair match to the 3.28 \micron\
feature (Fig.~\ref{fig:sgra_pah}).  However, this does not provide
simultaneously a good fit to the observed interstellar 6.2 \micron\ band
(cf., \S\ref{nightmare}, Fig.~\ref{fig:sgra_pah}).   In fact, as pointed
out by Schutte \etal\ (1998), neutral PAHs have too weak an absorption
feature to account for the observed strength  of the 6.2 \micron\ feature.
For that reason, Schutte \etal\ (1998) attributed the 6.2 \micron\ band to
ionized PAHs whose CC modes are intrinsically much stronger. Nevertheless,
this assignment still requires 10\% of the available carbon in the form of
PAH cations for GCS3 and 4.  

An origin of the 3.28 \micron\ band in HACS is also possible.  HACs, when
heated, lose much of their hydrogen and their aromatic component grows in
strength.   The spectral characteristics of heat-treated HACs (Scott \&
Duley 1996; Blanco \etal\ 1988) resemble those observed toward GCS3.  At
present, we cannot distinguish observationally whether free-flying,
ionized PAHs or HACs are
the carriers of the 3.28 \micron\ band in the diffuse ISM.

\section{Origin and Evolution of Aliphatic and Aromatic Hydrocarbon
Dust}\label{discussion}

Whereas the distribution of aliphatic hydrocarbon dust, detected via the
3.4, 6.85, and 7.25 \micron\ features, toward the GC shows little
variation, aromatic hydrocarbons identified via their C-H stretch (3.28
\micron) and C-C stretch (6.2 \micron), however, may not be evenly
distributed throughout the GC region (\S\ref{aromatics}).   It has
recently been proposed that the Quintuplet sources are dusty late-type WC
stars. WC stars are descendants of massive ($M>25$M$_{\odot}$) OB stars
where substantial mass loss and mixing have uncovered freshly synthesized
material at the stellar surface.  These objects are rich in He, C, and O
($\simeq$ 10:3:1 by number), but there is no evidence for H or N
($<10^{-2}$ and $10^{-3}$ relative to He by number;
\nocite{willis82,nugis82,torres88} Willis 1982; Nugis 1982; Torres 1988).
Thus, aromatic dust associated with the 6.2 \micron\ C-C stretch feature,
detected toward GCS3, GCS4, and the well-known WR stars, may instead be
associated with the environment of these objects instead of the general
diffuse ISM.  One of the IR sources (IRS6E) included in the ISO-SWS beam
centered on \sgra\ has been identified as a late-type WC star
\nocite{krabbe_etal95} (Krabbe \etal\ 1995);  this may explain the
presence of a small amount of 6.2 \micron\  absorption (relative to its
deep 9.7 \micron\ silicate feature).      However, while the WC stars may
be responsible for producing the carrier(s) of the C-C stretch (6.2
\micron),  they cannot be a significant contributor to the carrier(s) of
the  C-H stretch (3.28 \micron).  Presently, the connection between 3.28
and 6.2 \micron\ absorption features is not well established
observationally. 

Various models have been proposed for the origin and evolution of these
dust components -- ices, aliphatic, and aromatic dust -- in the ISM. Of
particular relevance are the processing of interstellar ices leading to
organic grains \nocite{greenberg89}(Greenberg 1989) and the processing of
Hydrogenated Amorphous Carbon \nocite{scott_duley96,blanco_etal88} (Scott
\& Duley 1996; Blanco \etal\ 1988). Ices are generally thought to form by
accretion on preexisting silicate cores inside dense molecular clouds.
These ices can be photolyzed by penetrating UV photons and may be
converted into an organic refractory mantle. Laboratory studies have shown
that this process may lead to a material with IR characteristics matching
the observed interstellar 3.4 \micron\ bands well
\nocite{greenberg_etal95,allamandola_sandford_valero88}(Greenberg \etal\
1995;  Allamandola, Sandford, \& Valero 1988). However, such residues
invariably show much stronger absorption in the 5-8 $\mu$m range due to
residual oxygen bonded to the carbon structures. Such bands are not
observed in the spectra of sources showing the 3.4 $\mu$m feature
(Figs.~\ref{fig:gcs3_6mic}, \ref{fig:sgra_diffuse}).  Our results thus
strengthen the argument (Chiar \etal\ 1998b; Adamson \etal\ 1999; see
\S\ref{intro}) that the 3.4 \micron\ carrier is not an organic refractory
mantle produced by irradiation of ices in molecular clouds.  

Alternatively, HAC, which in structure is akin to soot, could be an
important stardust component injected into the ISM by low mass stars
during the AGB phase of their evolution, in agreement with the strength of
the 3.4 \micron\ absorption band in the protoplanetary object CRL 618
(Chiar \etal\ 1998b). Extensive  analysis of  interstellar observations
require that the amount of C locked up in HAC relative to H-nuclei is
$\sim8\times10^{-5}$ (Furton \etal\ 1999).  This depends on the extent to
which the carbonaceous material is hydrogenated, but the adopted value
H/C$\sim0.5$ is most consistent with other HAC properties (Furton \etal\
1999). It has been suggested that HAC could also be formed through
accretionary processes in the diffuse ISM but plausible chemical
mechanisms have not been identified \nocite{duley_williams95}(Duley \&
Williams 1995). Nevertheless, HACs are thought to evolve in the diffuse
ISM: under the influence of FUV irradiation (or heating), HACs lose H and
their structure is transformed from an aliphatic structure into a more
aromatic structure \nocite{ogmen_duley88,furton_witt93} (Ogmen \& Duley
1988; Furton \& Witt 1993). This process will be counteracted by
hydrogenation.  In photodissociation regions, the astrophysical sites of
these processes, both UV irradiation and exposure to atomic hydrogen will
occur at the same time (Furton \& Witt 1993).  The relative amount of
aromatic and aliphatic carbon is then set by the balance between these two
processes and this will determine the relative strength of the 3.3 (and
6.2) \micron\ band compared to the 3.4 \micron\ band. We note here that,
if the 3.3 and 6.2 bands observed in GCS 3 are associated with the diffuse
ISM rather than circumstellar dust, we may actually be observing both the
aromatic and aliphatic components involved in this process along these
lines of sight.  However,  models of this type have problems too. In
particular, theoretical studies of dust destruction and observational
studies of elemental depletions suggest that the lifetime of stardust is
much shorter ($3\times10^8$ yr) than the injection timescale by old stars
($5\times 10^9$ yr), which suggests that  dust formation in the ISM is
very important \nocite{jones_etal94,jones_etal96}(Jones \etal\ 1994,
1996). Now, in the models, ices accreted inside dense molecular clouds,
would be rapidly destroyed when the molecular cloud is dispersed. In an
evolutionary sense, ices would have little influence on the dust
composition in the diffuse ISM.  However, it has been suggested that
organic residues produced by UV photolysis of interstellar ices are
rapidly carbonized (H, O, and N are lost) by the strong far-UV fields in
the diffuse ISM and that this leads to a HAC-like mantle
\nocite{jenniskens_etal93}(Jenniskens \etal\ 1993).  It remains puzzling
then that, while interstellar ice, as well as silicate, features can be
polarized, the 3.4 \micron\ aliphatic hydrocarbon feature is not (Adamson
\etal\ 1999).  Possibly, this might imply a link between the aliphatic
hydrocarbon carriers of the 3.4 \micron\ feature and the aromatic carrier
of the 2175\AA\ bump which is also not polarized
\nocite{martin_etal99}(Martin \etal\ 1999). In any case, a HAC-like
material with very little oxygen impurities, is the carrier of the 3.4,
6.85, and 7.25 \micron\ features. Further observations, particularly of
the 3.28, 3.4, and 6.2 $\mu$m bands will be instrumental in settling the
origin and evolution of this component.

\section{Summary}\label{summary}
We have identified molecular cloud and diffuse interstellar medium
material  along the line of sight to the Galactic Center using ISO-SWS
data for 3 sightlines.  The amount of molecular cloud material varies
across the GC field, and the amount of diffuse ISM material shows little
variation.  This is supported by our observations of the 3 \micron\
\water-ice feature which ranges between $\tau_{3.0}=0.16-0.50$, and the
3.4 \micron\ aliphatic hydrocarbon feature which ranges between
$\tau_{3.4}=0.15-0.21$.  The line of sight towards \sgra\ contains the
largest column of molecular cloud material, relative to GCS3 and 4,
apparent from the depth of the 3 \micron\ ice feature.  Towards all GC
sources discussed in the present paper, CO$_2$ ice is detected, and CO is
present mainly in the gas phase.   Towards \sgra,  solid \methane\
has also been firmly identified.  As the amount of solid CO$_2$ relative
to \water\ is higher toward the Quintuplet sources compared to \sgra,
there may be more than one molecular cloud associated with the overall GC
region.  However, the processes that control the solid CO$_2$ abundance
are not well understood \nocite{whittet_etal98}
(e.g., Whittet \etal\ 1998; Gerakines \etal\ 1999)

In many ways, the molecular clouds along the line of sight toward the
Galactic Center are similar to local molecular clouds in terms of
abundances of key solid state molecules (Table~6).  Water-ice remains the
most abundant solid state molecule; the solid CO$_2$/\water\ ratio is
within the range found for local molecular clouds (Gerakines \etal\ 1999);
\methanol/\water\ is similar to that observed in quiescent clouds
such as Taurus \nocite{chiar_adamson_whittet96}(Chiar \etal\ 1996), and
does not resemble the higher abundance observed toward heavily embedded
massive protostars.  While solid CO is not detected in the sightlines
discussed here, a weak 4.67 \micron\ feature has been detected previously
(McFadzean \etal\ 1989).  The CO/\water\ ratio implied is $<10\%$.  The 3.0
\micron\ ice profile is indicative of the temperature, as well as the
content of the ices.  The profile  observed toward the GC is distinct from
that seen in local molecular clouds such as Taurus.  It is  consistent
with cold ($\sim15$\,K) ices, similar to local quiescent dense clouds,
but   setting this feature apart from those studied in local clouds is
substructure at 2.95 \micron\ indicative of a relatively high abundance
(up to $\sim30\%$, relative to \water) of  \ammonia\ in the ices.  Toward
local molecular cloud sources, the solid \ammonia/\water\ abundance is
observed to be  $\sim10\%$ \nocite{whittet_etal96hh,lacy_etal98}(e.g.,
Whittet \etal\ 1996, Lacy \etal\ 1998).  The high ammonia abundance in GC
molecular clouds could reflect the higher N/O ratio in the inner Galaxy. 
 
Aliphatic hydrocarbons, first detected via their 3.4 \micron\ CH stretch
feature, are a widespread component of the diffuse interstellar medium.
ISO-SWS has for the first time detected the corresponding C-H deformation
modes at 6.85 and 7.25 \micron\ in the line of sight towards the Galactic
Center.  The ratio of integrated intensities for these three absorption
features are consistent with  the predicted values for saturated aliphatic
hydrocarbons \nocite{wexler67} (Wexler 1967).   Hydrogenated amorphous
carbon containing a substantial fraction of saturated aliphatic hydrocarbon
groups provides a convincing match to the observed hydrocarbon absorption
features, and is the likely carrier of these features.

Aromatic hydrocarbons, detected via their 3.3 \micron\ C-H and 6.2 \micron\
C-C stretching vibrations, do not appear to be evenly distributed
throughout the Galactic Center.    Absorption at 3.28 \micron\ is detected
in the line-of-sight toward GCS3, but not toward \sgra.  In addition,
the 6.2 \micron\ absorption feature toward \sgra\ is much weaker than that
toward the Quintuplet sources.  Although our results suggest that the
3.28 and 6.2 \micron\ features are due to the same carrier, it remains
unclear whether the C-C stretch absorption feature could instead be 
produced by dusty C-rich WR stars, and is thus independent of the
C-H stretch feature.

\nocite{chiar_etal98yso}
\nocite{schutte_etal99}
\nocite{tielens_etal91}
\nocite{dartois_etal99}
\nocite{moneti_cernicharo00}

\acknowledgements{J.E.C. was supported by a National Research Council
associateship for a portion of this work.  J.E.C. is currently supported by
NASA's Long Term Space Astrophysics program under grant 399-20-61-02.  D.C.B.W.
is supported by NASA through grants NAG5-7598, NAG5-7884, and JPL contract
961624.}

\newpage
%
%

%
%

%
%

%
%

%
%

%
%

%
%

%
%

%
%

%
%

%
%

%
%

%
%

%
%

\clearpage
\newpage
\begin{deluxetable}{lcccclc}
\tablecaption{Observational Details}
\tablewidth{0pt}
\tablecolumns{6}
\tablehead{
\colhead{Source} & Other ID & \multicolumn{2}{c}{Position (J2000)}   & \colhead{AOT\tablenotemark{a}} & \colhead{UTC date}  & \colhead{$t_{int}$} \\
 &                & \colhead{RA} & \colhead{Dec}           &  & & (sec)
}
\startdata
Sgr A* & &  17 45 40.0  & $-$29 00 29 & 1.4   &  1996 Feb. 19 & 6528    \\
Sgr A-W IRS3 & & 17 45 39.6 & $-$29 00 23 & 6   &  1997 Mar. 14 & 1086  \\ 

GCS3-I & MGM 5-4\tablenotemark{b} &  17 46 14.9  & $-$28 49 34  & 1.3   &  1996 Aug. 29 & 3454 \\
&         &              &              & 6     &  1996 Oct. 8  & 3226 \\

GCS4 & MGM 5-3\tablenotemark{b}  &  17 46 15.6  & $-$28 49 47  & 1.3   &  1996 Sept. 9 & 3454  \\

\enddata 
\tablenotetext{a}{AOT 1.m represents Astronomical Observing Template 1, 
speed m.}
\tablenotetext{b}{MGM identification from survey by Moneti, Glass, \& Moorwood 
1992}
\end{deluxetable}

\newpage

\begin{deluxetable}{lccccccccccc}
\rotate
\tablecolumns{12}
\tabletypesize{\small}
\tablewidth{0pc}
\tablenum{2}
\tablecaption{Optical Depths\tablenotemark{a}}
\tablehead{
\colhead{} & \multicolumn{4}{c}{Dense Cloud Dust Features} & \colhead{} & \multicolumn{6}{c}{Diffuse Cloud Dust Features} \\
\cline{2-5} \cline{7-12}
\colhead{Source} &  \colhead{$\tau_{3.0}$}  & \colhead{$\tau_{4.27}$} &
   \colhead{$\tau_{6.13}$\tablenotemark{b}} & \colhead{$\tau_{7.68}$} & \colhead{}   &
   \colhead{$\tau_{3.28}$} & \colhead{$\tau_{3.4}$}  & \colhead{$\tau_{6.2}$\tablenotemark{c}} &
   \colhead{$\tau_{6.85}$} & \colhead{$\tau_{7.25}$} & \colhead{$\tau_{9.7}$\tablenotemark{d}}
}
\startdata
Sgr A* & $0.50\pm0.01$ & $0.70\pm0.01$ & $0.13\pm0.02$ & 
   $0.017\pm0.003$ & & $<0.02$ & $0.21\pm0.01$  & $0.05\pm0.01$ &
   $0.05\pm0.01$ & $0.03\pm0.01$ & 3.52  \\
GCS3   & $0.23\pm0.02$ & $0.48\pm0.02$ & $0.03\pm0.01$ & $<0.02$   & 
       & $0.026\pm0.005$ & 
   $0.16\pm0.01$ & $0.10\pm0.02$ & $0.04\pm0.01$ & $<0.03$
   & 2.38 \\
GCS4   & $0.16\pm0.01$ & $0.22\pm0.02$ & $0.02\pm0.01$ & $<0.03$   
       &               & $<0.05$	  & 
   $0.15\pm0.01$  & $0.08\pm0.02$ & $<0.03$ & $<0.03$ & 2.24 \\
\enddata 
\tablenotetext{a}{Optical depth values for GCS3 are from AOT6 data where 
available; $\tau_{7.68}$ for Sgr\,A* is from AOT6 data; all others are determined from AOT1 spectra.}

\tablenotetext{b}{Optical depth is from the maximum allowable depth of the H$_2$O:NH$_3$:CO$_2$ mixture at 6.13 \micron,
which is constrained by the depth of the 3 \micron\ feature.  See \S4.1 for details.}

\tablenotetext{c}{Optical depth is that of the residual feature after subtracting the ice component.  See \S4.1
and \S5.2.1 for details.}

\tablenotetext{d}{Silicate optical depths from Schutte \etal\ 1996 and 
W.\ Vriend, D.\ Kester, \& A.\ Tielens, in preparation.}
\end{deluxetable}

\newpage
\begin{deluxetable}{lccccccc}
\tablenum{3}
\tablecaption{Ice Column Densities\tablenotemark{a}}
\tablewidth{0pt}
\tablecolumns{8}
\tablehead{
\colhead{Source} & \colhead{$N$(H$_2$O)} 
                 & \colhead{$N$(CO$_2$)\tablenotemark{b}} 
                 & \colhead{$N$(CO)} &
                   \colhead{$N$(CH$_4$)} &
                   \colhead{$N$(HCOOH)} &
                   \colhead{$N$(H$_2$CO)} &
                   \colhead{$N$(CH$_3$OH)} 
}
\startdata
Sgr A* & $12.4\pm2.5$  & $1.7\pm0.2$ & $<1.5$ & $0.30\pm0.07$ & $0.8\pm0.2$ 
       & $<0.3$ & $<0.5$ \\

GCS3   & $4.7\pm0.4$   & $1.1\pm0.1$ & $<0.9$ & $<0.3$ & \nodata\ 
       & \nodata\ & \nodata\ \\
            
GCS4   & $3.0\pm0.3$   & $0.7\pm0.2$ & $<0.9$ & $<0.5$ & \nodata\ 
       & \nodata\ & \nodata\ \\
\enddata 
\tablenotetext{a}{Column densities in units of $10^{17}$ cm$^{-2}$}
\tablenotetext{b}{Gerakines et al. 1999}
\end{deluxetable}

\begin{deluxetable}{ccc}
\tablenum{4}
\tablecaption{Aliphatic Hydrocarbon Bands: Sgr A*}
\tablewidth{0pt}
\tablecolumns{3}
\tablehead{
\colhead{Band (\micron)} & \colhead{$\tau(\lambda_0)$} 
    & \colhead{Integrated Area (cm$^{-1}$)}
}
\startdata
3.4  (SgrA*) & $0.21\pm0.01$ & 29  \\
6.85 (SgrA*) & $0.05\pm0.01$ & 1.6 \\
7.25 (SgrA*) & $0.03\pm0.01$ & 0.7 \\
\hline
\colhead{Relative Integrated Intensity} & \colhead{Observed} & \colhead{Expected\tablenotemark{a}} \\
\hline
A(6.85\micron)/A(7.25\micron) & 2.3   &  2.5 (branched), 5 (normal) \\
A(6.85\micron)/A(3.4\micron) & 0.06  &  0.07 \\
A(7.25\micron)/A(3.4\micron) & 0.02  &  0.03 (branched), 0.02 (normal) \\
\enddata 
\tablenotetext{a}{Expected integrated absorption area intensity for
saturated aliphatic hydrocarbons from Wexler 1967}
\end{deluxetable}

\begin{deluxetable}{cccc}
\tablenum{5}
\tablecaption{Aromatic Hydrocarbon Bands: GCS3\tablenotemark{a}}
\tablewidth{0pt}
\tablecolumns{4}
\tablehead{
\colhead{Band (\micron)} & \colhead{$\tau(\lambda_0)$} 
  & \colhead{FWHM (cm$^{-1}$)} & \colhead{Integrated Area (cm$^{-1}$)}
}
\startdata
3.28   & $0.026\pm0.005$ & $25\pm5$ &  $0.8\pm0.3$ \\
6.2    & $0.10\pm0.02$   & $34\pm5$ &  $3.6\pm0.6$ \\
\enddata 
\tablenotetext{a}{Uncertainty does not reflect uncertainties in continuum
choice}
\end{deluxetable}

\begin{deluxetable}{cccc}
\tablenum{6}
\tablecaption{Comparison of Ice Abundances toward Sagitarius A* and NGC
7538 IRS9\tablenotemark{a}}
\tablewidth{0pt}
\tablecolumns{4}
\tablehead{
\colhead{Molecule} & \colhead{SgrA*} & \colhead{NGC 7538 IRS9} & \colhead{Ref.\tablenotemark{b}}}
\startdata
H$_2$O     & 100   & 100 &  1 \\
NH$_3$     & $20-30$ & 10  &  2 \\
CO$_2$     & 14    & 20  &  3 \\
CO         & $<12$   & 15 &  4 \\
CH$_3$OH   & $<4$  &  7 &  5  \\
CH$_4$     &  2    & 2   &  6 \\
HCOOH      &  $6$    & 3   &  7 \\
\enddata
\tablenotetext{a}{Abundances are percent relative to \water-ice.}
\tablenotetext{b}{References are for NGC 7538 IRS9 ice column densities. 
Column densities for \sgra\ are from this paper.} \tablerefs{1.\ Tielens
\etal\ 1991, 2.\ Lacy \etal\ 1998, 3.\ Gerakines \etal\ 1999, 4.\ Chiar
\etal\ 1998a, 5.\ Dartois \etal\ 1999, 6.\ Boogert \etal\ 1998, 7.\
Schutte \etal\ 1999.}
\end{deluxetable}


\begin{thebibliography}{}

\bibitem[\protect\citeauthoryear{{Adamson} et~al.}{{Adamson}
  et~al.}{1999}]{adamson_etal99}
{Adamson}, A.~J., {Whittet}, D. C.~B., {Chrysostomou}, A., {Hough}, J.~H.,
  {Aitken}, D.~K., {Wright}, G.~S.,  \& {Roche}, P.~F. 1999, \apj, 512, 224

\bibitem[\protect\citeauthoryear{{Aitken} et~al.}{{Aitken}
  et~al.}{1986}]{aitken_etal86}
{Aitken}, D.~K., {Briggs}, G.~P., {Roche}, P.~F., {Bailey}, J.~A.,  \& {Hough},
  J.~H. 1986, \mnras, 218, 363

\bibitem[\protect\citeauthoryear{{Allamandola} et~al.}{{Allamandola}
  et~al.}{1992}]{allamandola_etal92}
{Allamandola}, L.~J., {Sandford}, S.~A., {Tielens}, A. G. G.~M.,  \& {Herbst},
  T.~M. 1992, \apj, 399, 134

\bibitem[\protect\citeauthoryear{{Allamandola}, {Sandford}, \&
  {Valero}}{{Allamandola} et~al.}{1988}]{allamandola_sandford_valero88}
{Allamandola}, L.~J., {Sandford}, S.~A.,  \& {Valero}, G.~J. 1988, Icarus, 76,
  225

\bibitem[\protect\citeauthoryear{{Becklin} et~al.}{{Becklin}
  et~al.}{1978}]{becklin_etal78I}
{Becklin}, E.~E., {Matthews}, K., {Neugebauer}, G.,  \& {Willner}, S.~P. 1978,
  \apj, 219, 121

\bibitem[\protect\citeauthoryear{{Blanco}, {Bussoletti}, \&
  {Colangeli}}{{Blanco} et~al.}{1988}]{blanco_etal88}
{Blanco}, A., {Bussoletti}, E.,  \& {Colangeli}, L. 1988, \apj, 334, 875

\bibitem[\protect\citeauthoryear{{Boogert} et~al.}{{Boogert}
  et~al.}{1998}]{boogert_etal98}
{Boogert}, A. C.~A., {Helmich}, F.~P., {van Dishoeck}, E.~F., {Schutte}, W.~A.,
  {Tielens}, A. G. G.~M.,  \& {Whittet}, D. C.~B. 1998, \aap, 336, 352

\bibitem[\protect\citeauthoryear{{Boogert} et~al.}{{Boogert}
  et~al.}{1997}]{boogert_etal97}
{Boogert}, A. C.~A., {Schutte}, W.~A., {Helmich}, F.~P., {Tielens}, A. G.
  G.~M.,  \& {Wooden}, D.~H. 1997, \aap, 317, 929

\bibitem[\protect\citeauthoryear{{Boogert} et~al.}{{Boogert}
  et~al.}{1996}]{boogert_etal96}
{Boogert}, A. C.~A., et~al. 1996, \aap, 315, L377

\bibitem[\protect\citeauthoryear{{Bridger}, {Wright}, \& {Geballe}}{{Bridger}
  et~al.}{1993}]{bridger_wright_geballe93}
{Bridger}, A., {Wright}, G.,  \& {Geballe}, T. 1993, in Conference Abstract
  Volume of Infrared Astronomy with Arrays: The Next Generation, ed.
  I.~{McLean}, UCLA Dept. of Astronomy, Los Angeles, 537

\bibitem[\protect\citeauthoryear{{Brooke}, {Sellgren}, \& {Geballe}}{{Brooke}
  et~al.}{1999}]{brooke_sellgren_geballe99}
{Brooke}, T.~Y., {Sellgren}, K.,  \& {Geballe}, T.~R. 1999, \apj, 517, 883

\bibitem[\protect\citeauthoryear{{Butchart} et~al.}{{Butchart}
  et~al.}{1986}]{butchart_etal86}
{Butchart}, I., {McFadzean}, A.~D., {Whittet}, D. C.~B., {Geballe}, T.~R.,  \&
  {Greenberg}, J.~M. 1986, \aap, 154, L5

\bibitem[\protect\citeauthoryear{{Chiar}, {Adamson}, \& {Whittet}}{{Chiar}
  et~al.}{1996}]{chiar_adamson_whittet96}
{Chiar}, J.~E., {Adamson}, A.~J.,  \& {Whittet}, D. C.~B. 1996, \apj, 472, 665

\bibitem[\protect\citeauthoryear{{Chiar} et~al.}{{Chiar}
  et~al.}{998a}]{chiar_etal98yso}
{Chiar}, J.~E., {Gerakines}, P.~A., {Whittet}, D. C.~B., {Pendleton}, Y.~J.,
  {Tielens}, A. G. G.~M.,  \& {Boogert}, A. C.~A. 1998a, \apj, 498, 716

\bibitem[\protect\citeauthoryear{{Chiar} et~al.}{{Chiar}
  et~al.}{998b}]{chiar_etal98crl}
{Chiar}, J.~E., {Pendleton}, Y.~J., {Geballe}, T.~G.,  \& {Tielens}, A. G.
  G.~M. 1998b, \apj, 507, 281

\bibitem[\protect\citeauthoryear{{Dartois} et~al.}{{Dartois}
  et~al.}{1999}]{dartois_etal99}
{Dartois}, E., {Schutte}, W., {Geballe}, T.~R., {Demyk}, K., {Ehrenfreund}, P.,
   \& {d'Hendecourt}, L. 1999, \aap, 342, L32

\bibitem[\protect\citeauthoryear{{de Graauw} et~al.}{{de Graauw}
  et~al.}{996a}]{degraauw_etal96a}
{de Graauw}, T., et~al. 1996a, \aap, 315, L49

\bibitem[\protect\citeauthoryear{{de Graauw} et~al.}{{de Graauw}
  et~al.}{996b}]{degraauw_etal96b}
{de Graauw}, T., et~al. 1996b, \aap, 315, L345

\bibitem[\protect\citeauthoryear{{d'Hendecourt} et~al.}{{d'Hendecourt}
  et~al.}{1996}]{dhendecourt_etal96}
{d'Hendecourt}, L., et~al. 1996, \aap, 315, L365

\bibitem[\protect\citeauthoryear{{Duley} et~al.}{{Duley}
  et~al.}{1998}]{duley_etal98}
{Duley}, W.~W., {Scott}, A.~D., {Seahra}, S.,  \& {Dadswell}, G. 1998, \apjl,
  503, L183

\bibitem[\protect\citeauthoryear{{Duley} \& {Williams}}{{Duley} \&
  {Williams}}{1995}]{duley_williams95}
{Duley}, W.~W.,  \& {Williams}, D.~A. 1995, \mnras, 272, 442

\bibitem[\protect\citeauthoryear{{Eckart} et~al.}{{Eckart}
  et~al.}{1995}]{eckart_etal95}
{Eckart}, A., {Genzel}, R., {Hofmann}, R., {Sams}, B.~J.,  \& {Tacconi-Garman},
  L.~E. 1995, \apjl, 445, L23

\bibitem[\protect\citeauthoryear{{Figer}, {McLean}, \& {Morris}}{{Figer}
  et~al.}{1999}]{figer_mclean_morris99}
{Figer}, D.~F., {McLean}, I.~S.,  \& {Morris}, M. 1999, \apj, 514, 202

\bibitem[\protect\citeauthoryear{{Furton}, {Laiho}, \& {Witt}}{{Furton}
  et~al.}{1999}]{furton_etal99}
{Furton}, D.~G., {Laiho}, J.~W.,  \& {Witt}, A.~N. 1999, \apj, 526, 752

\bibitem[\protect\citeauthoryear{{Furton} \& {Witt}}{{Furton} \&
  {Witt}}{1993}]{furton_witt93}
{Furton}, D.~G.,  \& {Witt}, A.~N. 1993, \apjl, 415, L51

\bibitem[\protect\citeauthoryear{{Gerakines}, {Schutte}, \&
  {Ehrenfreund}}{{Gerakines} et~al.}{1996}]{gerakines_schutte_ehrenfreund96}
{Gerakines}, P.~A., {Schutte}, W.~A.,  \& {Ehrenfreund}, P. 1996, \aap, 312,
  289

\bibitem[\protect\citeauthoryear{{Gerakines} et~al.}{{Gerakines}
  et~al.}{1995}]{gerakines_etal95}
{Gerakines}, P.~A., {Schutte}, W.~A., {Greenberg}, J.~M.,  \& {van Dishoeck},
  E.~F. 1995, \aap, 296, 810

\bibitem[\protect\citeauthoryear{{Gerakines} et~al.}{{Gerakines}
  et~al.}{1999}]{gerakines_etal99}
{Gerakines}, P.~A., et~al. 1999, \apj, 522, 357

\bibitem[\protect\citeauthoryear{{Giard} et~al.}{{Giard}
  et~al.}{1988}]{giard_etal88}
{Giard}, M., {Serra}, G., {Caux}, E., {Pajot}, F.,  \& {Lamarre}, J.~M. 1988,
  \aap, 201, L1

\bibitem[\protect\citeauthoryear{{Gibb} et~al.}{{Gibb}
  et~al.}{2000}]{gibb_etal00}
{Gibb}, E.~L., et~al. 2000, \apj, in press

\bibitem[\protect\citeauthoryear{{Greenberg}}{{Greenberg}}{1989}]{greenberg89}
{Greenberg}, J.~M. 1989, in {Interstellar Dust: {IAU} Symposium no. 135}, ed.
  L.~J. {Allamandola} \& A.~G. G.~M. {Tielens} (Dordrecht: Reidel), 345

\bibitem[\protect\citeauthoryear{{Greenberg} et~al.}{{Greenberg}
  et~al.}{1995}]{greenberg_etal95}
{Greenberg}, J.~M., {Li}, A., {Mendoza-Gomez}, C.~X., {Schutte}, W.~A.,
  {Gerakines}, P.~A.,  \& {De Groot}, M. 1995, \apjl, 455, L177

\bibitem[\protect\citeauthoryear{{Hagen} \& {Tielens}}{{Hagen} \&
  {Tielens}}{1981}]{hagen_tielens_greenberg81}
{Hagen}, W.,  \& {Tielens}, J.~M., A. G. G. M.~{Greenberg}. 1981, Chem.~Phys.,
  56

\bibitem[\protect\citeauthoryear{{Hudgins} \& {Sandford}}{{Hudgins} \&
  {Sandford}}{998a}]{hudgins_sandford98a}
{Hudgins}, D.~M.,  \& {Sandford}, S.~A. 1998a, J.~Phys.~Chem.~A, 102, 329

\bibitem[\protect\citeauthoryear{{Hudgins} \& {Sandford}}{{Hudgins} \&
  {Sandford}}{998b}]{hudgins_sandford98b}
{Hudgins}, D.~M.,  \& {Sandford}, S.~A. 1998b, J.~Phys.~Chem.~A, 102, 344

\bibitem[\protect\citeauthoryear{{Hudgins} et~al.}{{Hudgins}
  et~al.}{1993}]{hudgins_etal93}
{Hudgins}, D.~M., {Sandford}, S.~A., {Allamandola}, L.~J.,  \& {Tielens}, A. G.
  G.~M. 1993, \apjs, 86, 713

\bibitem[\protect\citeauthoryear{{Jenniskens} et~al.}{{Jenniskens}
  et~al.}{1993}]{jenniskens_etal93}
{Jenniskens}, P., {Baratta}, G.~A., {Kouchi}, A., {De Groot}, M.~S.,
  {Greenberg}, J.~M.,  \& {Strazzulla}, G. 1993, \aap, 273, 583

\bibitem[\protect\citeauthoryear{{Joblin} et~al.}{{Joblin}
  et~al.}{1995}]{joblin_etal95}
{Joblin}, C., {Boissel}, P., {Leger}, A., {D'Hendecourt}, L.,  \& {Defourneau},
  D. 1995, \aap, 299, 835

\bibitem[\protect\citeauthoryear{{Jones}, {Tielens}, \& {Hollenbach}}{{Jones}
  et~al.}{1996}]{jones_etal96}
{Jones}, A.~P., {Tielens}, A. G. G.~M.,  \& {Hollenbach}, D.~J. 1996, \apj,
  469, 740

\bibitem[\protect\citeauthoryear{{Jones} et~al.}{{Jones}
  et~al.}{1994}]{jones_etal94}
{Jones}, A.~P., {Tielens}, A. G. G.~M., {Hollenbach}, D.~J.,  \& {MCKee}, C.~F.
  1994, \apj, 433, 797

\bibitem[\protect\citeauthoryear{{Keane} et~al.}{{Keane}
  et~al.}{2000}]{keane_etal00}
{Keane}, J.~V., {Tielens}, A. G. G.~M., {Boogert}, A. C.~A., {Schutte}, W.~A.,
  \& {Whittet}, D. C.~B. 2000, \aap, submitted

\bibitem[\protect\citeauthoryear{{Kobayashi} et~al.}{{Kobayashi}
  et~al.}{1983}]{kobayashi_etal83}
{Kobayashi}, Y., {Okuda}, H., {Sato}, S., {Jugaku}, J.,  \& {Dyck}, H.~M. 1983,
  \pasj, 35, 101

\bibitem[\protect\citeauthoryear{{Krabbe} et~al.}{{Krabbe}
  et~al.}{1995}]{krabbe_etal95}
{Krabbe}, A., et~al. 1995, \apjl, 447, L95

\bibitem[\protect\citeauthoryear{{Lacy} et~al.}{{Lacy}
  et~al.}{1998}]{lacy_etal98}
{Lacy}, J.~H., {Faraji}, H., {Sandford}, S.~A.,  \& {Allamandola}, L.~J. 1998,
  \apjl, 501, L105

\bibitem[\protect\citeauthoryear{{Lebofsky}, {Rieke}, \& {Tokunaga}}{{Lebofsky}
  et~al.}{1982}]{lebofsky_etal82}
{Lebofsky}, M.~J., {Rieke}, G.~H.,  \& {Tokunaga}, A.~T. 1982, \apj, 263, 736

\bibitem[\protect\citeauthoryear{{Lequeux} \& {Jourdain de Muizon}}{{Lequeux}
  \& {Jourdain de Muizon}}{1990}]{lequeux_muizon90}
{Lequeux}, J.,  \& {Jourdain de Muizon}, M. 1990, \aap, 240, L19

\bibitem[\protect\citeauthoryear{{Lutz} et~al.}{{Lutz}
  et~al.}{1996}]{lutz_etal96}
{Lutz}, D., et~al. 1996, \aap, 315, L269

\bibitem[\protect\citeauthoryear{{Mar\'echal}}{{Mar\'echal}}{1987}]{marechal87}
{Mar\'echal}, Y. 1987, \jcp, 87, 6344

\bibitem[\protect\citeauthoryear{{Martin}, {Clayton}, \& {Wolff}}{{Martin}
  et~al.}{1999}]{martin_etal99}
{Martin}, P.~G., {Clayton}, G.~C.,  \& {Wolff}, M.~J. 1999, \apj, 510, 905

\bibitem[\protect\citeauthoryear{{Mattila} et~al.}{{Mattila}
  et~al.}{1996}]{mattila_etal96}
{Mattila}, K., {Lemke}, D., {Haikala}, L.~K., {Laureijs}, R.~J., {Leger}, A.,
  {Lehtinen}, K., {Leinert}, C.,  \& {Mezger}, P.~G. 1996, \aap, 315, L353

\bibitem[\protect\citeauthoryear{{McFadzean} et~al.}{{McFadzean}
  et~al.}{1989}]{mcfadzean_etal89}
{McFadzean}, A.~D., {Whittet}, D. C.~B., {Bode}, M.~F., {Adamson}, A.~J.,  \&
  {Longmore}, A.~J. 1989, \mnras, 241, 873

\bibitem[\protect\citeauthoryear{{McKee}}{{McKee}}{1989}]{mckee89}
{McKee}, C.~F. 1989, \apj, 345, 782

\bibitem[\protect\citeauthoryear{{Moneti} \& {Cernicharo}}{{Moneti} \&
  {Cernicharo}}{2000}]{moneti_cernicharo00}
{Moneti}, A.,  \& {Cernicharo}, J. 2000, in ISO Beyond the Peaks: The 2nd ISO
  workshop on analytical spectroscopy, E61

\bibitem[\protect\citeauthoryear{{Moneti}, {Glass}, \& {Moorwood}}{{Moneti}
  et~al.}{1992}]{moneti_etal92}
{Moneti}, A., {Glass}, I.,  \& {Moorwood}, A. 1992, \mnras, 258, 705

\bibitem[\protect\citeauthoryear{{Moneti}, {Glass}, \& {Moorwood}}{{Moneti}
  et~al.}{1994}]{moneti_etal94}
{Moneti}, A., {Glass}, I.~S.,  \& {Moorwood}, A. F.~M. 1994, \mnras, 268, 194

\bibitem[\protect\citeauthoryear{{Moore}, {Ferrante}, \& {Nuth}}{{Moore}
  et~al.}{1996}]{moore_ferrante_nuth96}
{Moore}, M.~H., {Ferrante}, R.~F.,  \& {Nuth}, J. A.~I. 1996, \planss, 44, 927

\bibitem[\protect\citeauthoryear{{Nagata} et~al.}{{Nagata}
  et~al.}{1990}]{nagata_etal90}
{Nagata}, T., {Woodward}, C.~E., {Shure}, M., {Pipher}, J.~L.,  \& {Okuda}, H.
  1990, \apj, 351, 83

\bibitem[\protect\citeauthoryear{{Nugis}}{{Nugis}}{1982}]{nugis82}
{Nugis}, T. 1982, in Wolf Rayet stars: Observations, Physics, Evolution, ed.
  C.~W.~H. {de Loore} \& A.~J. {Willis} (Dordrecht: Kluwer), 131

\bibitem[\protect\citeauthoryear{{Ogmen} \& {Duley}}{{Ogmen} \&
  {Duley}}{1988}]{ogmen_duley88}
{Ogmen}, M.,  \& {Duley}, W.~W. 1988, \apjl, 334, L117

\bibitem[\protect\citeauthoryear{{Okuda} et~al.}{{Okuda}
  et~al.}{1990}]{okuda_etal90}
{Okuda}, H., et~al. 1990, \apj, 351, 89

\bibitem[\protect\citeauthoryear{{Pendleton} et~al.}{{Pendleton}
  et~al.}{1994}]{pendleton_etal94}
{Pendleton}, Y.~J., {Sandford}, S.~A., {Allamandola}, L.~J., {Tielens}, A. G.
  G.~M.,  \& {Sellgren}, K. 1994, \apj, 437, 683

\bibitem[\protect\citeauthoryear{{Pendleton}, {Tielens}, \&
  {Werner}}{{Pendleton} et~al.}{1990}]{pendleton_tielens_werner90}
{Pendleton}, Y.~J., {Tielens}, A. G. G.~M.,  \& {Werner}, M.~W. 1990, \apj,
  349, 107

\bibitem[\protect\citeauthoryear{{Rieke}, {Rieke}, \& {Paul}}{{Rieke}
  et~al.}{1989}]{rieke_etal89}
{Rieke}, G.~H., {Rieke}, M.~J.,  \& {Paul}, A.~E. 1989, \apj, 336, 752

\bibitem[\protect\citeauthoryear{{Ristorcelli} et~al.}{{Ristorcelli}
  et~al.}{1994}]{ristorcelli_etal94}
{Ristorcelli}, I., {Giard}, M., {Meny}, C., {Serra}, G., {Lamarre}, J.~M., {Le
  Naour}, C., {Leotin}, J.,  \& {Pajot}, F. 1994, \aap, 286, L23

\bibitem[\protect\citeauthoryear{{Roche} \& {Aitken}}{{Roche} \&
  {Aitken}}{1985}]{roche_aitken85}
{Roche}, P.~F.,  \& {Aitken}, D.~K. 1985, \mnras, 215, 425

\bibitem[\protect\citeauthoryear{{Rubin} et~al.}{{Rubin}
  et~al.}{1988}]{rubin_etal88}
{Rubin}, R.~H., {Simpson}, J.~P., {Erickson}, E.~F.,  \& {Haas}, M.~R. 1988,
  \apj, 327, 377

\bibitem[\protect\citeauthoryear{{Sandford} et~al.}{{Sandford}
  et~al.}{1991}]{sandford_etal91}
{Sandford}, S.~A., {Allamandola}, L.~J., {Tielens}, A., {Sellgren}, K.,
  {Tapia}, M.,  \& {Pendleton}, Y. 1991, \apj, 371, 607

\bibitem[\protect\citeauthoryear{{Schutte} et~al.}{{Schutte}
  et~al.}{1999}]{schutte_etal99}
{Schutte}, W.~A., et~al. 1999, \aap, 343, 966

\bibitem[\protect\citeauthoryear{{Schutte} et~al.}{{Schutte}
  et~al.}{1996a}]{schutte_etal96a}
{Schutte}, W.~A., {Gerakines}, P.~A., {Geballe}, T.~R., {van Dishoeck}, E.~F.,
  \& {Greenberg}, J.~M. 1996a, \aap, 309, 633

\bibitem[\protect\citeauthoryear{{Schutte} \& {Greenberg}}{{Schutte} \&
  {Greenberg}}{1989}]{schutte_greenberg89}
{Schutte}, W.~A.,  \& {Greenberg}, J.~M. 1989, in Dust in the universe, ed.
  M.~E. {Bailey} \& D.~A. {Williams} (Cambridge and New York: Cambridge
  University Press), 403

\bibitem[\protect\citeauthoryear{{Schutte} et~al.}{{Schutte}
  et~al.}{1996b}]{schutte_etal96b}
{Schutte}, W.~A., et~al. 1996b, \aap, 315, L333

\bibitem[\protect\citeauthoryear{{Schutte} et~al.}{{Schutte}
  et~al.}{1998}]{schutte_etal98}
{Schutte}, W.~A., et~al. 1998, \aap, 337, 261

\bibitem[\protect\citeauthoryear{{Scott} \& {Duley}}{{Scott} \&
  {Duley}}{1996}]{scott_duley96}
{Scott}, A.,  \& {Duley}, W.~W. 1996, \apjl, 472, L123

\bibitem[\protect\citeauthoryear{{Sellgren} et~al.}{{Sellgren}
  et~al.}{1995}]{sellgren_etal95}
{Sellgren}, K., {Brooke}, T.~Y., {Smith}, R.~G.,  \& {Geballe}, T.~R. 1995,
  \apjl, 449, L69

\bibitem[\protect\citeauthoryear{{Sellgren} et~al.}{{Sellgren}
  et~al.}{1987}]{sellgren_etal87}
{Sellgren}, K., {Hall}, D. N.~B., {Kleinmann}, S.~G.,  \& {Scoville}, N.~Z.
  1987, \apj, 317, 881

\bibitem[\protect\citeauthoryear{{Simpson} et~al.}{{Simpson}
  et~al.}{1995}]{simpson_etal95}
{Simpson}, J.~P., {Colgan}, S. W.~J., {Rubin}, R.~H., {Erickson}, E.~F.,  \&
  {Haas}, M.~R. 1995, \apj, 444, 721

\bibitem[\protect\citeauthoryear{{Skinner} et~al.}{{Skinner}
  et~al.}{1992}]{skinner_etal92}
{Skinner}, C.~J., {Tielens}, A. G. G.~M., {Barlow}, M.~J.,  \& {Justtanont}, K.
  1992, \apjl, 399, L79

\bibitem[\protect\citeauthoryear{{Smith}, {Sellgren}, \& {Tokunaga}}{{Smith}
  et~al.}{1989}]{smith_sellgren_tokunaga89}
{Smith}, R.~G., {Sellgren}, K.,  \& {Tokunaga}, A.~T. 1989, \apj, 344, 413

\bibitem[\protect\citeauthoryear{{Strazzulla}, {Castorina}, \&
  {Palumbo}}{{Strazzulla} et~al.}{1995}]{strazzulla_castornia_palumbo95}
{Strazzulla}, G., {Castorina}, A.~C.,  \& {Palumbo}, M.~E. 1995, \planss, 43,
  1247

\bibitem[\protect\citeauthoryear{{Tielens} \& {Allamandola}}{{Tielens} \&
  {Allamandola}}{1987}]{tielens_allamandola87b}
{Tielens}, A. G. G.~M.,  \& {Allamandola}, L.~J. 1987, in Interstellar
  Processes, ed. D.~{Hollenbach} \& H.~{Thronson} (Dordrecht: Kluwer), 397

\bibitem[\protect\citeauthoryear{{Tielens} et~al.}{{Tielens}
  et~al.}{1991}]{tielens_etal91}
{Tielens}, A. G. G.~M., {Tokunaga}, A.~T., {Geballe}, T.~R.,  \& {Baas}, F.
  1991, \apj, 381, 181

\bibitem[\protect\citeauthoryear{{Tielens} et~al.}{{Tielens}
  et~al.}{1996}]{tielens_etal96}
{Tielens}, A. G. G.~M., {Wooden}, D.~H., {Allamandola}, L.~J., {Bregman}, J.,
  \& {Witteborn}, F.~C. 1996, \apj, 461, 210

\bibitem[\protect\citeauthoryear{{Torres}}{{Torres}}{1988}]{torres88}
{Torres}, A.~V. 1988, \apj, 325, 759

\bibitem[\protect\citeauthoryear{{van der Hucht} et~al.}{{van der Hucht}
  et~al.}{1996}]{vanderhucht_etal96}
{van der Hucht}, K.~A., et~al. 1996, \aap, 315, L193

\bibitem[\protect\citeauthoryear{{Wada}, {Sakata}, \& {Tokunaga}}{{Wada}
  et~al.}{1991}]{wada_sakata_tokunaga91}
{Wada}, S., {Sakata}, A.,  \& {Tokunaga}, A.~T. 1991, \apjl, 375, L17

\bibitem[\protect\citeauthoryear{{Wexler}}{{Wexler}}{1967}]{wexler67}
{Wexler}, A.~S. 1967, Appl.\ Spectrosc.\ Rev., 1, 29

\bibitem[\protect\citeauthoryear{{Whittet} et~al.}{{Whittet}
  et~al.}{1997}]{whittet_etal97}
{Whittet}, D. C.~B., et~al. 1997, \apj, 490, 729

\bibitem[\protect\citeauthoryear{{Whittet} et~al.}{{Whittet}
  et~al.}{1998}]{whittet_etal98}
{Whittet}, D. C.~B., et~al. 1998, \apjl, 498, L159

\bibitem[\protect\citeauthoryear{{Whittet} et~al.}{{Whittet}
  et~al.}{1996}]{whittet_etal96hh}
{Whittet}, D. C.~B., et~al. 1996, \apj, 458, 363

\bibitem[\protect\citeauthoryear{{Willis}}{{Willis}}{1982}]{willis82}
{Willis}, A.~J. 1982, in Wolf Rayet stars: Observations, Physics, Evolution,
  ed. C.~W.~H. {de Loore} \& A.~J. {Willis} (Dordrecht: Kluwer), 87

\bibitem[\protect\citeauthoryear{{Willner} et~al.}{{Willner}
  et~al.}{1979}]{willner_etal79}
{Willner}, S.~P., {Russell}, R.~W., {Puetter}, R.~C., {Soifer}, B.~T.,  \&
  {Harvey}, P.~M. 1979, \apjl, 229, L65

\bibitem[\protect\citeauthoryear{{Wollman}, {Smith}, \& {Larson}}{{Wollman}
  et~al.}{1982}]{wollman_smith_larson82}
{Wollman}, E.~R., {Smith}, H.~A.,  \& {Larson}, H.~P. 1982, \apj, 258, 506

\end{thebibliography}
\end{document}